# An Automatable Analytical Algorithm for Structure-Based Protein Functional Annotation *via* Detection of Specific Ligand 3D Binding Sites

## *Application to ATP Binding Sites in ser/thr Protein Kinases and GTP Binding Sites in Small Ras-type G-Proteins*

**Vicente M. Reyes, Ph.D.***
**E-mail: vmrsbi.RIT.biology@gmail.com**

*work done at:

Dept. of Pharmacology, School of Medicine,
University of California, San Diego
9500 Gilman Drive, La Jolla, CA 92093-0636
&
Dept. of Biological Sciences, School of Life Sciences
Rochester Institute of Technology
One Lomb Memorial Drive, Rochester, NY 14623

Running title: Prediction of Specific Ligand Binding Sites in Proteins

**Abstract.** We have developed an analytical, ligand-specific and scalable algorithm that detects a 'signature' of the 3D binding site of a given ligand in a protein 3D structure. The said signature is a 3D motif in the form of an irregular tetrahedron whose vertices represent the backbone or side-chain centroids of the amino acid residues at the binding site that physically interact with the bound ligand atoms. The motif is determined from a set of solved training structures, all of which bind the ligand. Just as alignment of linear amino acid sequences enables one to determine consensus sequences in proteins, the present method allows the determination of three-dimensional consensus structures or 'motifs' in folded proteins. Although such is accomplished by the present method not by alignment of 3D protein structures or parts thereof (e.g., alignment of ligand atoms from different structures) but by nearest-neighbor analysis of ligand atoms in protein-bound forms, the same effect, and thus the same goal, is achieved. We have applied our method to the prediction of GTP- and ATP-binding protein families, namely, the small Ras-type G-protein and ser/thr protein kinase families. Validation tests reveal that the specificity of our method is nearly 100% for both protein families, and a sensitivity of ≥ 60% for the ser/thr protein kinase family and approx. 93% for the small, Ras-type G-protein family. Further tests reveal that our algorithm can distinguish effectively between GTP and GTP-like ligands, and between ATP- and ATP-like ligands. The method was applied to a set of predicted (by 123D threading) protein structures from the slime mold (*D. dictyostelium*) proteome, with promising results.

## 1 Introduction

*Motivation and Background.* The last 10-15 years saw genome sequencing efforts around the world give rise to a deluge of genetic sequence information, a significant percentage of which code for novel proteins of unknown function (Bentley & Parkhill, 2004; Murphy et al., 2004; Holm & Sander, 1996; Baxevanis, 2003; Miller et al., 2004). Currently there are approximately 1,500 gene sequences deposited in GenBank per protein structure deposited in the PDB; by mid-2010, this imbalance is projected to reach a rate of over 5,000 gene sequences in GenBank per structure in the PDB, even though structural genomics projects have been steadily producing experimental structures of proteins with novel folds and unknown functions (Berman & Westbrook, 2004; Yakunin et al., 2004; Jung & Lee, 2004; Norvell & Machalek, 2000; Burley, 2000; Terwilliger, 2000; Heinemann, 2000; Yokoyama et al., 2000). This work addresses the need for structure-based approaches to protein function prediction by (a.) the assignment of putative functions for these proteins based on their solved structures, and (b.) the prediction of function of proteins which are unsolved but which have acceptably accurate 3D structure models predicted from their sequences; in this case, the function is predicted from the 3D models. Our approach to this problem is to devise an analytical and automatable (hence amenable to high-throughput implementation) procedure that can predict which ligand binding site a given protein harbors, given its 3D structure. We apply the method to the prediction of ATP- and GTP-binding proteins.

*Overview of the Methodology.* The basic underlying idea behind our method is shown in Figure 1. The amino acid residues (R1, R2, R3 and R4) contacting the ligand at its binding site are generally far apart in primary sequence and occur in no particular order. They interact with specific ligand atoms either *via* hydrogen bonding or van der Waals interactions, and either *via* side-chain or backbone. The identities of the amino acids are conserved to varying degrees, but in general they are more conserved if the interaction with ligand is *via* side-chain than if *via* backbone. Their relative distances are also generally conserved, as these are dictated by the ligand's molecular dimensions (which are constant), and bound conformation (often, but not always, conserved).

Our overall methodology may be deemed as a three-stage procedure (see Figure 2), namely, determination of the tertrahedral 'three-dimensional search motif' (3D SM) from the training set (part I), validation of the 3D SM using the positive and negative control structures (part II), and the application of the 3D SM search algorithm to proteins of unknown function , i.e., with unknown ligand binding sites (part III) .

In part I, the 'binding site consensus motif' - the collection of physical interactions, namely, H-bonds and VDW interactions, between the ligand and its binding site in the receptor protein - of the ligand of interest is first determined from a set of training structures. The training protein structures are first transformed into the 'double-centroid reduced representation' (DCRR). The 3D SM is then derived from the consensus motif. The relation between the usual all-atom representation of a protein (as in the PDB) and its corresponding DCRR is shown in Figure 3. In DCRR, each amino acid residue of the protein is represented by two points: the centroid of its backbone atoms (N, CA, C', O), and the centroid of its side-chain atoms (CB, CG, etc.), reducing the atomicity of the protein by >76%.

The resulting 3D SM is shown in Figure 4. The 3D SM is an irregular tetrahedron whose four vertices are the centroids of the backbone and/or side-chain atoms of the residues making the major interactions with the ligand in its binding site (ligand binding site, LBS).

The 3D SM is next validated in par t II using positive and negative control structures. These are, respectively, experimentally solved protein structures which are known to be able to bind, and be unable to bind, the ligand of interest. ROC curves and then plotted to assess the sensitivity and specificity of the procedure using the 3D SM.

Part III is simply the actual application of the method to proteins of unknown function. This involves the screening for the presence of the 3D SM in the test set: the set of experimental or predicted structures of proteins whose functions are being sought, the functional assignment being inferred from the presence or absence of the 3D SM. To implement the algorithm, the four vertices of the tetrahedral 3D SM must be in 'tree' data structure, i.e., exactly one (arbitrarily selected) must be designated as the 'root', and the other three, 'node1', 'node2' and 'node3', designated as 'n1', 'n2' and 'n3', respectively (sometimes designated as 'e1', 'e2' and 'e3', respectively, as well). Our screening algorithm is written in Fortran77 and 90, and requires the parameters of the 3D SM, as input (i.e., the identities of the amino acid residues making up the 3D SM, the lengths of its 6 sides, and the nature of their interaction with the ligand, i.e., *via* backbone or *via* side chain, a total of at least 14 parameters). Candidate 3D SM 'sides' are sequestered by the algorithm from the test protein structure 3D structure based on the input parameters as well as the 'connectivity' (i.e., whether each node is connected to the same root, as well as to each other). In ascertaining connectivity, groups of centroids called' clusters' are first selected, from which groups called 'trees' are further selected. An 'error margin', ε, typically ±1.4 Å, is added to the lengths of the sides of the 3D SM to incorporate a fuzzy element into the screening process.

## 2 Methods, Procedures and Datasets

*Selection of Training Sets.* As both GTP- and ATP-binding proteins are both very heterogeneous, we decided to focus on the small, Ras-tywe G-protein and the ser/thr protein kinase families. The training structures for the GTP-binding small, Ras-type G-protein family are shown and described in Table 1, Panel A. All are small Ras-type G-proteins except 1LOO, which, interestingly, is mouse muscle adenylosuccinate synthetase. The training structures for ATP-binding ser/thr protein kinase family are shown and described in Table 1, Panel B. All are CDK2, except for 1GOl, which is a MAPK, and 1PHK, 2PHK and 1QL6, which are phosphorylase kinases. All training structures have bound GTP or ATP, and not analogs, since analogs are known sometimes to bind in a novel fashion with the receptor protein (as exemplified by dihydrofolate reductase with bound folate versus methotrexate, see Matthews et al., 1977; Bystroff et al., 1990).

*Determining the 3D Ligand Binding Site Consensus Motif.* All calculations and operations were done using Fortran 77 or 90 programs and UNIX scripts. Nearest neighbors of each ligand atom in thet raining proteins in all-atom representation (AAR) are first determined (see Figure 5). The H-bonding and VDW interactions were then selected from this set of nearest neighbors. These are tabulated for clarity and ease of analysis by inspection (see Figure 6). Those interactions occurring in all or most of the training structures, and/or those with ideal distances between interacting atoms, were selected. The resulting selection is the 3D binding site consensus motif (3D BS CM) for the specific ligand in the protein family under consideration. The H-bonding and VDW interactions in the 3D LBS CM are used to guide the selection of backbone and side chain centroids for inclusion in the 3D SM. The determination of the 3D ligand binding site consensus motif is described in more detail in our previous work (Reyes, V.M. & Sheth, V.N., 2011).

*Building Reduced Representations of Protein Structures in the Training Set.* The protein structures were transformed into DCRR, where each residue is represented by two points: the centroid of the backbone atoms (N, CA, C', O), and the centroid of the side chain atoms (CB, CG, etc.). DCRR reduces the atomicity of the protein by >76%, making the operations more economical without losing too much chemical information. The conversion of the all-atom PDB representation of the protein into DCRR is described in more detail in our previous work (Reyes, V.M. & Sheth, V.N., 2011).

*Determining the 3D SM.* The 3D SM is derived from the 3D LBS CM. The protein file is transformed into DCRR then the nearest neighbors of each ligand atom in the protein are determined. Then the root and nodes of the 3D SM are selected from these nearest neighbor centroids by cross-comparison with the 3D LBS CM (in AAR) determined earlier, based on prevalence and agreement of lengths (H-bonds and VDW interactions) to ideal values (Bondi, et al., J. Phys. Chem., 68:441, 1964).

*Screening for the 3D SM.* The screening procedure for the 3D SM has been described schematically in our previous publication (Reyes, V.M. & Sheth, V.N., 2011). The overall screening algorithm is applied using the 3D SM, which is in double-centroid representation (DCRR; see Figure 7, Panel A) and is composed of nine steps (see Figure 7, Panel B), all of which are automated except the last, which is done by inspection. Each of the nine steps is implemented as a Fortran 77/90 program; all are incorporated in a script, allowing the user to perform all nine steps at once, but the last.

*Validation Tests.* To estimate the sensitivity and specificity of the algorithm for each protein family, positive and negative controls were performed. The positive control structures (n=15) for the GTP-binding small, Ras-type G-protein family are: 1C1Y, 1GWN, 1NVX, 1PLK, 1QRA, 1ZBD, 3RAP, 521P, 1AS3, 1BOF, 1GIT, 1KAO, 1PLL, 1TAG and 2RAP. They are described in Table 2, Panel A. The positive control structures (n=15) for the ATP-binding ser/thr protein kinase family are: 1CDK, 1FMO, 1GY3, 1JBP, 1Q24, 1S9I, 1S9J, 1UA2, 2CPK, 1CSN, 1H1W, 1L3R, 1OGU, 1RDQ and 1CM8. They are described in Table 2, Panel B. The negative control structures (n=30) used for all protein families are: 135L, 1A1M, 1A6T, 1BHC, 1PSN, 1BRF, 1EWK, 1CBN, 1MV5, 1JFF, 104M, 1ASH, 1B3B, 1BRF, 1CKO, 1CRP, 1EWK, 1F3O, 1FW5, 1HWY, 1JBP, 1MJJ, 1MV5, 1NQT, 1OGU, 1PE6, 1RDQ, 1SVS,

1TWY and 1Z3C. They are described in Table 3. The screening algorithm was performed on both control sets with no distinction between tight and loose binding sites. Specificity (Sp), sensitivity (Sn), success rate (SR) and the Matthews correlation coefficient (MCC) were calculated as follows: Sp = TN/(TN + FP); Sn = TP/(TP + FN); SR = (TP + TN)/(TP + TN + FP + FN); MCC = (TP*TN - FP*FN)/sqrt[(TP + FP)*(TP + FN)*(TN + FP)*(TN + FN)] ; where TP = true positives; TN = true negatives; FP = false positives; and FN = false negatives.

## 3 Results

### Determining the 3D Binding Site Consensus Motif for GTP.

*Hydrogen-Bonding Interactions.* Based on the protein-ligand H-bonding and VDW interactions common among the training structures, the 3D binding site consensus motif (3D BS CM) is determined. The H-bonding interactions that each GTP atom makes with the amino acid residues at the binding site in the receptor protein in each of the training structures are first tabulated (data not shown). The most common H-bonding interactions shared by members of the training set involve pyrimidine N1 and ring substituents atoms N2 and O6 of GTP. In all cases, pyrimidine N1 and ring substituent N2 atoms are respectively engaged in ideal H-bonds to atoms OE1 and OE2 of an asp residue, while pyrimidine substituent atom O6 is H-bonded to the backbone N of either a lys (one structure) or ala residue (the rest). Interestingly, GTP atoms N1, N2 and O6 are precisely the ones involved in base-pairing interactions involving the guanylate moiety in ribonucleic acids (RNA). Pyrimidine N3 and imidazole nitrogens N7 and N9 do not participate in any H-bonding interactions in all cases. Meanwhile, in structures E, F and P, ribose hydroxyl oxygens O2* and O3* are engaged in ideal H-bonds with backbone oxygens of a his, val or ile residues, and those of a glu or asp residues, respectively, while in structure B, the same ribose oxygens are simultaneously H-bonded to the backbone O of a gly residue. There are no H- bonding interactions involving ribose O2* and O3* in structures M, N and O. Note that we did not consider the binding cavity of the triphosphate moiety of either GTP or ATP because: (1) pyrophosphate sometimes binds proteins nonspecifically; (2) phosphate is a common moiety in biological ligands, and, most importantly; and (3) the triphosphate tail has a high degree of rotational freedom with respect to the guanosine and adenosine moieties (*syn* and *anti* conformations), therefore a search motif containing a node or root that corresponds to a triphosphate atom is unlikely to have branches and node-edges that have fixed lengths.

*Van der Waals Interactions.* As in the previous section, the van der Waals interactions that each GTP atom makes with the amino acid residues at the binding site in the receptor protein in each of the training structures are tabulated (data not shown). For simplicity, VDW interactions involving non-C protein atoms (N, O, P and S) are not considered if they are already involved in H-bonding interaction/s; this approximation is reasonable as, all other things being equal, a H-bonding interaction is about 10 times stronger than a VDW interaction. The most prevalent protein-GTP VDW interactions in family 02B involve three amino acid residues: a lys, a phe and a tyr, but is most pronounced in structures B, F, N and P. Specifically, ribose C1*, and pyrimidine C5 and C6 are all within VDW interacting distance with CE and CG atoms of a lys, while ribose C2* and pyrimidine C4 both lie close to the CZ atom of a phe; finally, ribose C3* and C4* are both close to the CB atom of a tyr. In structure E, ribose C4* and C5* are sufficiently close to atom CB of a pro; in structure O, ribose C2* and imidazole C8 lie close to the CB atom of a thr, as well as CA of a gly in the other*;* and finally in structure M, ribose C2* and imidazole C8 lie close to the CD and CG atoms, respectively, of two lys residues different from the lys referred to above.

In summary, H-bonding and VDW interactions between GTP and protein prevalent in structures in this family are presented in Figure 8, Panel A. We note that the binding mode of GTP in this family is characterized by polar interactions between ligand and protein involving the extremities of the guanosine moiety: pyrimidine N1 and substituents N2 and O6 on one end, and ribose O2* and O3* on the other. It as well features non-polar interactions between ligand and protein involving the central portion of the moiety, including the distal edge of the ribose ring composed of atoms C3* and C4*. The four polar atoms in the central part of guanosine – pyrimidine N3, imidazole N7 and N9, and ribose O4* - are not engaged in H-bonding with protein.

### Determining the 3D Binding Site Motif Consensus for ATP.

*Hydrogen-Bonding Interactions.* The procedure was similar to that described for the GTP-binding family the previous section. The most common interactions between ATP and protein are those that involve pyrimidine N1 and pyrimidine substituent N6 of the ligand. In all cases pyrimidine N1 is H-bonded either to a leu or a met backbone N, while pyrimidine N6 is H-bonded to either an asp or a glu backbone O. There is a further pattern: interaction of N1 with a leu and of N6 with an asp are highly correlated; so does the interaction of N1 with a met and N6 with a glu. Interestingly, pyrimidine N1 and N6 are precisely the ones involved in base-pairing interactions involving the adenylate moiety in RNA. On the other hand, pyrimidine N3 and imidazole nitrogens N7 and N9, do not participate in any H-bond interactions. The two ribose hydroxyl oxygens O2* and O3* are also involved in several interactions, notably (a.) between the backbone O of a gln residue and either ribose O2* (structures A, B and E) or O3* (structures G2, I, J1 and J2); (b.) between the OD2 atom of an asp and either ribose O2* (structures D, J1 and J2) or O3* (structure E); (c.) between the OE2 atom of a glu and either ribose O2* (structures I and K) or O3* (structure H); and (d.) between the NZ atom of a lys and ribose O3* (structure D).

*Van der Waals Interactions.* Again, the procedure was similar to that described for the GTP-binding family the previous section. The most common protein-ATP VDW interactions in members of the ser/thr PK training set are those between (a.) pyrimidine C6 and either the CB atom of an ala or the CD1 atom of a leu; (b.) imidazole C8 and either the CG1 atom of a val or the CD1 atom of a leu, and (c.) ribose C4* and the CA atom of a gly. They occur in all 13 training structures except the last interaction, which is not present in structure D. We note that there are significantly more protein-ATP carbon-carbon VDW interactions in this training set than protein-GTP VDW interactions in the previous 2 training sets.

Figure 8, panel B summarizes the protein-ATP interactions in this training set. We note that binding of ATP is characterized by polar interactions along the extremities of the adenosine moiety, notably involving pyrimidine N1 and N6 on one end and ribose O2* and O3* on the other. Meanwhile, the central part of the moiety is characterized by non-polar interactions, notably ribose C3* and C4*. Of interest is the fact that four polar atoms in the central part of the adenosine moiety – pyrimidine N3, imidazole N7 and N9, and ribose O4* - are not involved in H-bonding.

**Deriving the 3D SM for GTP.**

*Preliminary Considerations.* The 3D SM (which is in DCRR) may be considered to be the the distillation of the physical interactions included in the 3D BS CM (which is in ARR), and the 3D SM is essentially a signature of the ligand binding site. During screening, the ligand is conceptually in AAR while the protein is in DCRR. Such a ligand environment composed of centroids does not represent genuine physical interaction, and for clarity we designate it as "*association*", and reserve the term "*interaction*" for actual physical interactions as H-bonding and VDW attractive forces. In constructing the 3D SM, the four most prevalent and/or ideal *interactions* in the training set are selected, and the corresponding centroids are then taken as elements of the 3D SM; these are the protein-ligand *associations*. In most cases, the best associations are obvious, and it is clear which centroids must be included in the 3D SM. Sometimes, however, there is a 'tie'. In such cases, the root or one or more nodes in the 3D SM is/are not unique and the researcher may incorporate a disjunction ("or") in the 3D SM. Since the protein is in DCRR in the 3D SM, we shall adopt the notation "X(b)" and "Z(s)" to denote "the backbone centroid of amino acid X" and "the side-chain centroid of amino acid Z", respectively. We denote a disjunction by a slash, e.g., L(s)/I(s) means 'leucine side chain or isoleucine side chain'

*Association Between Protein Backbone Centroids and Ligand Atoms.* In all the training structures for this family, ribose O5* is associated with a G(b), while pyrimidine O6 is associated with an A(b) or a G(b). Similarly, pyrimidine N2 is associated with a K(b), and imidazole C8 is associated with a G(b) - the same G(b) associated with ribose O5*. Comparing the above associations against the 3D BS CM, found earlier, we note that ribose O5* is indeed H-bonded to the backbone N of a gly residue in all cases except structure M, while pyrimidine O6 is H-bonded to the backbone N of an ala residue, except again for structure M where it is H-bonded to the backbone N of a lys, and as well to backbones N and O of two different gly residues. In contrast, pyrimidine N2 is not involved in H-bonding, and there is only weak VDW interaction between imidazole C8 and a gly residue. These findings suggest that G(b) and A(b)/G(b), associated respectively with ribose O5* and pyrimidine O6, may be included in the 3D search motif.

*Association Between Protein Side-Chain Centroids and Ligand Atoms.* In all the training structures for this family, pyrimidine N1 and ribose O4* are associated with a D(s) and a K(s), respectively. In addition, pyrimidine N3 and C4 are both associated with the same K(s) above. Comparing these associations against the 3D BC CM determined earlier, we note that in all cases except structure E, pyrimidine N1 indeed forms a H-bond with the OD1 and OD2 atoms of an asp residue, while ribose O4* is H-bonded the NZ atom of a lys. Pyrimidine N3, on the other hand, does not participate in any interactions except in structure M, where it is H-bonded to the backbone O of a gly and perhaps as well to the backbone N of a lys. The above information suggest that D(s) and K(s) may be included in the 3D search motif.

Combining the above information with the ones from the previous section, we come up with the 3D SM for this family, see Figure 9, Panel A. The 3D SM is composed of D(s) as root, A(b)/G(b) as node1, K(s) as node2, and G(b) as node3, which are associated with pyrimidine N1, pyrimidine O6, ribose O4*, and ribose O5*, respectively, of GTP. The lengths of the three branches (Rn1, Rn2 and Rn3) and three node-edges (n1n2, n1n3 and n2n3) complete the 3D SM; each is the averages from all the training structures. In most cases, the standard deviations are less than 1.0 Å, the rest being not much greater than 1.0 Å.

**Deriving the 3D SM for ATP.**

*Association Between Protein Backbone Centroids and Ligand Atoms.* In all the training structures for this family, we readily observe the following associations: (a.) pyrimidine N6 with a E(b) or an D(b), (b.) pyrimidine N1 with a L(b) or a M(b), (c.) imidazole N7 with an A(b), (d.) ribose O3* with either a E(B) or a Q(b) (except for structure G2), (e.) ribose O4* with a G(b), and (f.) pyrimidine C2 and C6 both with a L(b) or a M(b). Comparing the above associations against the 3D BS CM for this family found earlier, indeed we find that pyrimidine N6 is H-bonded to the backbone O of either a glu or an asp, validating (a.) above. Similarly, pyrimidine N1 is H-bonded to the backbone N of a leu or a met, thus validating (b.) above. In contrast, imidazole N7 is not involved in H-bonding, so (c.) above cannot be validated. The case for ribose O3* is not strong, as it is not H-bonded to a glu or a gln backbone O or N in 5 of the 13 training structures; we thus rule out (d.) as well. Cases (e.) and (f.) above are similarly weak. These findings suggest that E(b)/D(b) and L(b)/M(b), associated respectively with pyrimidine N6 and N1, may be included in the 3D search motif.

*Association Between Protein Side-Chain Centroids and Ligand Atoms*. In all the training structures for this family, we readily observe the following associations: (a.) ribose O2* with an D(s) or a E(s), (b.) pyrimidine N1 and N6 both with an A(s), (c.) ribose O4* with either a V(s) or a G(s), (d.) imidazole C8 with a V(s), and (e.) pyrimidine C5 and C6 both with an A(s). Again, comparing the above associations against the 3D BS CM for this family found earlier, we note that ribose O2* is indeed H-bonded to the OD2 or the OE2 atom of an asp or a glu, respectively, in all cases except structures C2 and G2, validating association (a.) above. In contrast, neither pyrimidine N1 nor N6 is H-bonded with an ala in any of the 13 training structures, ruling out (b.) above. Meanwhile, ribose O4* is in H-bonding interaction in only 3 of the 13 training structures, and never with a val; thus (c.) above may be eliminated. Except for structure D, imidazole C8 is always in VDW interaction with a val, thus validating (d.) above. Finally, pyrimidine C5 and C6 are only weakly involved in VDW interaction with an ala, hence (e.) is eliminated. The above findings indicate that we may include D(s)/E(s) and V(s), which are associated with ribose O2* and imidazole C8 respectively, in the 3D search motif for this family.

Combining the above results with those from the previous section yields the four centroids comprising the 3D SM for this family, see Figure 9, Panel B. The 3D SM is composed of E(b)/D(b) as root, V(s) as node1, D(s)/E(s) as node2 and L(b)/M(b) as node3, which are associated with pyrimidine N6 and N1, ribose O2* and imidazole C8, respectively, of ATP. The lengths of the three branches (Rn1, Rn2 and Rn3) and three node-edges (n1n2, n1n3 and n2n3) complete the 3D SM; as before, each is the averages from all the training structures. In most cases, the standard deviations are less than 1.0 Å, the rest being not much greater than 1.0 Å

**The Control Experiments.**

Ideally, the positive control structures should be structures which are known to contain the particular 3D SM; similarly, negative controls must be structures which are known to ***not*** contain the 3D SM. Since there is no practical way of determining the existence of the 3D SM's in protein 3D structures except by visual examination, we simply chose positive controls from the PDB by their similarity to the training structures, and negative controls by their dissimilarity with the training structures. Thus the results of the control tests presented below must be interpreted with caution.

*Validation Tests for the Small Ras-type G-protein Family.* Fourteen out of the 15 positive controls tested positive under the screening algorithm, while all 30 negative controls tested negative. Thus there are 14 true positives, 1 false negative, 30 true negatives, and 0 false positive. This suggests that, for this family, our algorithm has a sensitivity of 93.3%, a specificity of 100.0%, a success rate of 97.8%, and a Matthews correlation coefficient of 95.0%.

*Validation Tests for the ser/thr PK Family.* Nine out of the 15 positive controls tested positive under our algorithm, while all of the 30 negative controls tested negative. Thus there are 9 true positives, 6 false negatives, 30 true negatives, and 0 false positives. This suggests that, for this family, our algorithm has a sensitivity of 60.0%, a specificity of 100.0%, a success rate of 86.7%, and a Matthews correlation coefficient of 70.7%.

*ROC Curves.* Partial ROC curves constructed from the above results using the fact that ROC curves are parabolas that pass through the points (0,0) and (1,1) are shown in Figure 10, Panels A (left) and B (right) respectively, for the GTP-binding small, Ras-type G-proteins and the ser/thr protein kinase families.

**Expanded Test Sets for the GTP- and ATP- binding families.**

To further ascertain the sensitivity of our algorithm to the detection of the GTP- and ATP-binding sites in the two protein families, we tested more positive structures for both families using our screening algorithm. The expanded test set (positive structures, n=45) for the Ras-type GP family are: 1A2B, 1A2K, 1CTQ, 1FZQ, 1GG2, 1GP2, 1GUA, 1HUQ, 1JAH, 1JAI, 1K5D, 1K5G, 1KAO, 1MH1, 1N6L, 1NVU, 1NVV, 1NVW, 1OIV, 1OIW, 1OIX, 1QRA, 1R2Q, 1R4A, 1RYF, 1RYH, 1T91, 1WA5, 1WQ1, 1YHN, 1YZT, 1YZU, 1Z08, 1Z0A, 1Z0D, 1Z0J, 1ZBD, 1ZC3, 1ZC4, 2BKU, 2EW1, 2RAP, 3RAB, 3RAP and 5P21. They were selected from the PDB by keyword search, "Ras-type G-protein." Using the algorithm to screen them for the 3D SM of the GTP-binding site in the small, Ras-type G-proteins, all tested positive. These results indicate that the algorithm has high sensitivity for this family (see Figure 11, Panel A).

The expanded test set (positive structures, n=31) for the ATP-binding ser/thr PK family are: 1BO1, 1IA9, 1E8X, 1CJA, 1NW1, 1J7U, 1CDK, 1O6L, 1OMW, 1H1W, 1MUO, 1TKI, 1JKL, 1A06, 1PHK, 1KWP, 1IA8, 1GNG, 1HCK, 1JNK, 1HOW, 1LP4, 1F3M, 1O6Y, 1CSN, 1B6C, 2SRC, 1LUF, 1IR3, 1M14 and 1GJO. They are the 31 representative protein kinase structures used in the paper by Scheeff and Bourne (PLOS CB, 2005). Using the algorithm to screen them for the 3D SM of the ATP-binding site in the ser/thr protein kinase family, all but two tested positive. The two which tested negative were 1CJA and 1NW1, which were classified by Scheeff & Bourne to be "atypical protein kinases" due to the atypical architecture of their ATP binding sites. These results indicate that the algorithm has high sensitivity for this family (see Figure 11, Panel B).

**Additional Tests to Assess Discriminatory Power of Algorithm.**

*Screening of Proteins Binding GTP-like Ligands:* To ascertain the ability of our method to distinguish between GTP and GTP-like ligands, the following proteins that bind GTP-like ligands were randomly selected from the PDB: a GDP-binding protein (1kv3), a GMP-binding protein (1znx) and a cyclic GMP-binding protein (1q3e). Each were then screened using the search motif derived from the small Ras-type G-protein family (02B), initially with $\varepsilon$= 1.00 Å. All three structures tested negative, being eliminated early on in the screening process. A second screening with a more relaxed $\varepsilon$= 1.40 Å similarly produced negative results from all three structures. These results suggest that the ligand binding sites of these proteins are quite different from that of family 02B, and that our algorithm can effectively differentiate between them.

*Screening of Proteins Binding ATP-like Ligands:* To ascertain the ability of our method to distinguish between ATP and ATP-like ligands, the following proteins that bind ATP-like ligands were randomly selected from the PDB: a SAH-binding protein (10mh), a NAD-binding protein (1axe) an ADP-(and FAD-) binding protein (1cnf), an FAD-binding protein (1jrx), an AMP-(and FAD-) binding protein (1t9g), and a cAMP-binding protein (1ykd). Each were then screened using the search motif derived from the ser/thr PK protein family (01a), initially with $\varepsilon$= 1.00 Å. Five of the six structures yielded negative results, while structure 1jrx gave a positive result. Upon closer inspection, however, the putative ATP-binding site signature detected by the algorithm in structure 1jrx (namely, Asp-513, Val-539, Asp-358 and Met-511) turned out to be a spurious motif that is about 15 Å away from the genuine FAD binding site. This motif, although apparently resembling the ser/thr PK search motif, most probably does not have biological significance and occurred by chance. A second screening with a more relaxed $\varepsilon$= 1.40 Å again produced negative results from the same 5 of the 6 structures as before, but picked up a new putative ATP binding site signature in structure 1jrx. Upon closer inspection, this second ATP-binding site (namely, Glu-156, Val-294, Asp-125 and Leu-

154) was indeed located approximately where the adenosine moiety of FAD is bound in the protein (data not shown). However, the interaction of the residues in the search motif with the FAD adenosine moiety was only partial: that is, although carboxyl atoms OE1 and OE2 of Glu-156 (the root) are ideally H-bonded to the ribose hydroxyls AO3* and AO2* of the FAD adenosine moiety (at 2.67 Å and 2.51 Å, respectively), atom CG1 of Val-294 (node 1) is in VDW interaction not with FAD adenosine AC8 but with AN1, and the interaction is weak (at 6.97 Å); atom C of Leu-154 (node 3) is in VDW interaction not with FAD adenosine AN1 but with AC2, and the interaction is even weaker (at 7.74 Å); and Asp-125 (node 2) is not in H-bonding at all with any FAD adenosine atom. Taken together, these results demonstrate that our algorithm can effectively differentiate the ATP-binding site of the ser/thr PK family from the ATP-like ligand binding sites of the above proteins.

*Comparison of Method with Global Structure-Based Methods.* To demonstrate that the present method, which is local structure-based, works where global structure-based methods do not, consider the set of ATP-binding proteins with differing global folds but with similar ATP-binding site architectures, compiled by Kobayashi & Go (1997a; 1997b). Using the 4 ATP-containing structures in the set - 1cdk, 2dln, 1csn and 1gsa - for training, the ATP-binding site consensus motif determined from them yields the following 3D search motif: {root = L(b)/V(b); node1 = D(b)/K(b)/E(b)/N(b); node2 = L(s)/I(s)/V(s); node3 = D(b)/D(s)/E(b)/E(s); length Rn1 = 6.152; length Rn2 = 7.731; length Rn3 = 10.334; length n1n2 = 7.765; length n1n3 = 13.894; and length n2n3 = 10.279; with $\varepsilon = 1.00$ Å}. This search motif was used to screen the other 8 structures in the group - 2glt, 1scub, 1bnc, 1irk, 1dik, 1gtr, 1ses and 1lgr. All 8 tested positive (data not shown). This confirms that our algorithm can pick up all of these structures despite the fact that they have differing overall global folds. When the four training structures were inputted into the fold-based method, DALI, only training structures 2dln and 1gsa were able to pick up structures in the test set: 1bnc and 1dik. Even so, the ranks were low (2dln picked up 1bnc with rank 5, and 1dik with rank 12; 1gsa picked up 1bnc with rank 7, and 1dik with rank 14) and the RMSD's were high (2dln picked up 1bnc with RMSD of 6.0Å, and 1dik with RMSD of 3.1Å; 1gsa picked up 1bnc with RMSD of 5.4Å, and 1dik with RMSD of 3.7Å). The other two training structures (1cdk and 1csn) did not pick up any structure from the test set. Thus six of the eight test structures were invisible to DALI. The above results demonstrate that the present algorithm, being local-structure-based, is able to assign function to proteins where global fold-based methods such as DALI fail.

*Benchmark Experiment: Use of Predicted Structures.* We applied our method to the newly-sequenced proteome of the slime mold, *Dictyostelium discoideum* (Eichinger et al., 2005). A subset of the slime mold proteome was used to build 3D structures models of proteins from primary sequence using the threading program, 123D (Alexandrov et al., 1995), and the side chain modeling and partial refinement program, Modeller 6.2 (Sali et al., 1993). We selected proteins which did not have functional annotation at the time. The resulting protein models, 400 in all, were screened using the 3D SM for the two protein families. 47 structures tested positive for the Ras-type G-protein family, of which 10 have 'tight' binding sites ($\varepsilon \leq 1.0$ Å), and 37 have 'loose' binding sites (1.0 Å $\leq \varepsilon \leq$ 2.0 Å). Meanwhile, 52 structures tested positive for the ser/thr protein kinase family, of which 46 have 'tight' binding sites, and 6 have 'loose' binding sites (data not shown). These sets of structures have common elements. Preliminary independent results support the above findings. Most notably, a *Dictyostelium* protein containing the FNIP domain (marked by phe-asn-ile-phe repeats), which is of unknown function and occurring exclusively in amoeboid proteomes, has been found to possess the GTP binding site using our algorithm. Thus, our method seems to hold promise for function prediction using *predicted* structures, instead of experimentally solved structures.

## 4 Discussion

**Salient Features of the Method.** In the following sections, we discuss the salient features of our method and try to relate them to existing methods that have the same objective of predicting functional sites in protein 3D structures.

*Ability to Detect Consensus Binding Sites in Proteins of Different Folds*. From the foregoing sections, it is obvious that the present method achieves the detection of three-dimensional motifs in the binding sites of 3D protein structures by nearest- neighbor analysis of the bound ligand, and not by C-α alignment of protein structures. Such is an important consideration because it is known that proteins of different folds – which are not C-α alignable – can and do bind the same ligand using the same 3D binding site consensus. This is one strength of the present method.

*Inherent Objectivity.* The 3D SM in our method is based purely on protein-ligand H-bond and VDW interactions - information that is completely derivable from the PDB file. Similar existing methods involve construction of 'templates' based on catalytic site residues, determined with unavoidable subjective input from the experimenter and derived by human perusal of the literature. The fact that our approach is based on binding residues instead of catalytic residues means it is more objective than other similar methods.

*Ligand Specificity.* Other existing methods of functional site prediction involves merely specifying local sections of the protein 3D structural surface that are biologically important, without providing information exactly what the local site does – e.g., is it a catalytic site? is it a ligand binding site, and if so, what specific ligand does it bind? is it involved in protein-protein interaction, and if so, what is the partner protein?. In contrast, our method is ligand-specific: the researcher decides on the specific ligand for study, then collects a set of training structures from the PDB containing that ligand. Such ligand-specificity makes functional assignment more straightforward and objective.

*Extensibility and Scalability.* Our method is amenable to automation and thus well-suited for large-scale function assignment to entire proteomes. Other methods involve molecular dynamics simulations, and hence not easily automatable. Here, our method is applied to ATP and GTP binding site prediction, but it is also applicable to other ligands. Recently we have used it to predict sialic acid, retinoic acid, and heme-bound and unbound nitric oxide binding sites in experimentally solved protein structures in the PDB that currently do not have functional annotation (Reyes, V.M., unpublished [a.]). We have also extended the method to the prediction of specific protein-protein interaction partners (Reyes, V.M., unpublished [b.]). These applications demonstrate the extensibility and the versatility of the method. Additionally, the present method can be applied to non-protein structures, such as carbohydrate and lipid structures, which do not possess "standard" monomers (e.g., amino acids).

*Analytical Nature.* Other methods of protein functional site prediction make use of traditional machine learning techniques such as SVM, etc. Being statistical in nature, these methods do not allow the researcher full control over the classification process and involves some inexactitude in the results. In contrast, our algorithm is analytical, and as such it is one of only a handful of deterministic methods of functional site prediction currently available.

*Local Structure Dependence.* Our algorithm is designed to detect protein 'local structures' - the constellation of amino acid residues that come together in 3D space when the protein folds to assume its 3D structure (see Figure 1). The 3D SM defined in this work is precisely such a constellation of residues in space; our algorithm detects this spatial arrangement of residues in the input structure. A key concept here is that proper folding of the protein into its correct 3D structure is a *prerequisite* to the correct assembly of local structure.

*Use of Novel 'Tree' Data Structure.* To implement our algorithm, the 3D SM must be in 'tree' data structure, composed of a unique root, and typically 3 nodes, which together form an irregular tetrahedron. The qualitative and quantitative input parameters of the algorithm are all embodied in the 3D SM - the amino acid identities of the root and nodes, their interaction type (side chain or backbone), and the lengths of the 6 edges of the tetrahedron. The lengths of the 6 edges are determined by direct calculation from the atomic coordinates. Thus the optimal input parameter values are essentially dictated by the ligand's molecular dimensions. Although in theory the search motif can have 2 or more than 3 nodes, pilot tests demonstrate that a search motif composed of exactly 4 centroids (1 root and 3 nodes) has sufficient sensitivity and specificity to detect genuine ligand binding sites. Including even one more node results in a steep increase in computation time but little improvement in accuracy. Thus in this work we use search motifs composed of exactly 1 root and 3 nodes.

*Consideration of H-Bond and VDW Interactions Between Protein and Ligand.* Our method takes into account both H-bonds and VDW interactions in building the 3D SM. Other methods consider only polar interactions (i.e., H-bonds) between ligand and protein. However, ligand stabilization at the active site does not depend solely on H-bonding interactions; VDW attractive forces also play a significant role. Although a typical H-bond (2 - 10 kcal/mol) is about 10 times stronger than a typical VDW attractive force (0.1 to 1.1 kcal/mol), in our protein-ligand systems (ligand = ATP or GTP), there are approximately 8-9 VDW attractive forces per H-bond. It may thus be concluded that the contributions of H-bonding and VDW attractive force to ligand stabilization at the binding site are roughly of the same order of magnitude. This is most likely to be true in general.

*Use of Double Centroid Residue Reduced Representation.* Other methods use CA atoms to represent amino acid residues; this is not ideal, as it results in significant loss of information (see next section). Our method makes use of the 'double centroid residue representation' during screening for the ligand binding site. In DCRR, each protein residue is represented by two points: the centroid of its backbone atoms (N, CA, C' O), and the centroid of its side chain atoms (CB, CG, etc.,). Although the protein's atomicity is reduced by >76% compared to >88% using CA atoms, it is more than compensated for by the greater resolution and accuracy achieved (see below). A DCRR-like representation has been used before (Kleywegt, 1999), but unbeknownst to the author during the early stages of this study. Use of reduced protein representations such as DCRR also adds a fuzzy element into the screening process, thus counteracting the inherent uncertainty in screening predicted structures. Even in instances where the input structures are experimental or predicted with high accuracy, use of DCRR minimizes uncertainties resulting from protein dynamics and flexibility.

*Consideration of Both Protein Side Chain and Backbone Interactions.* Just as consideration of protein global fold alone is insufficient in predicting protein function, consideration of primary sequence alone is also ineffective. First, to consider sequence alone means to neglect protein folding altogether. During protein folding, often residues that are nonconsecutive and/or separated by any number of residues in the primary sequence come together in 3D space to form the local structure. Second, the backbone N-H and/or C=O of one or more binding residues are commonly involved in H-bonding with certain ligand atoms. This enables their corresponding side chains to *almost* freely assume any identity, thus allowing a high degree of sequence variation at these sites. Our method gives equal consideration to both backbone-ligand and side chain-ligand interactions. About 40% of all protein-ligand H-bonds in the training structures involve the protein N or O backbone atoms, while the rest involve side-chain functional groups; it is thus clear that protein backbone-ligand and protein side chain-ligand interactions both contribute significantly to the stabilization of the bound ligand.

*Use of Fuzzy Additive Term.* Our method incorporates am error margin, $\pm\varepsilon$, as an element of "fuzziness" to the screening process that is meant to minimize and/or counteract the inherent uncertainty in predicted structures and/or counteract the effects of protein flexibility.

*Semi-Docking Nature of Method.* Our method may be considered a semi-docking procedure. This is because not only is the precise geometry of the 3D SM known, but also the identity of its root and nodes. Thus the algorithm can search for the specific amino acids and their mode of interaction with the ligand (*via* backbone or side chain). Similarly, the specific ligand atoms interacting with the root and nodes are also precisely known. Thus the precise orientation of the ligand relative to its binding site in the protein is precisely determined. This is an advantage over existing methods which simply identify the location of functional sites.

**Limitations of the Method.**

First, we have currently no provision to automatically (computationally) ascertain the relative location of the detected 3D SM in the protein, i.e., whether it is located on the surface, lodged in a deep crevice, or close to the protein centroid. To this end, we have recently developed two complementary algorithms to quantify the degree of burial of ligand binding sites in proteins called the 'cutting plane' and 'tangent sphere' methods (V. M. Reyes, unpublished [c.]). Second, the error margin, $\varepsilon$, incorporated into the branch and node-edge lengths is usually in the order of 1.0 – 1.5 Å. Thus in cases where the protein assumes drastic conformational changes upon ligand binding, our method might fail. Third, in the determination of H-bonds between protein and ligand, we do not ascertain the linearity of the bonds of the interacting atoms - we merely measure non-hydrogen interatomic distances. However this issue is unlikely to have a significant effect on our results, as we sequester only those with perfect or near-perfect H-bond distances (2.7Å-2.9 Å). Cases in which the H-bonding atoms have perfect or near-prefect H-bonding distances *and* at the same time non-linear, are quite rare.

**Acknowledgments**
This work was supported by a fellowship award to V.M.R. from the Institutional Research and Academic Career Development Award (NIH GM 68524) through the Dept. of Pharmacology, School of Medicine at the University of California-San Diego (UCSD). The author also acknowledges the various forms of support and services kindly provided to him by the San Diego Supercomputer Center at the UCSD campus, the Academic Computing Services of the UCSD, the Division of Research Computing at the Rochester Institute of Technology (RIT), and the computing resources from the Dept. of Biological Sciences, College of Science, at RIT. .

**Figure Legends:**

**Figure 1.** ***Preservation of Protein Local Structure Regardless of Global Structure.*** The basic underlying idea behind much of the rationale for the work presented here is shown. The ligand binding site, a local structure, is formed by the coming together in 3D space of residues usually far apart and in no particular sequence in the protein primary sequence. The ligand binding residues may be conserved to varying degrees, as the interaction with ligand may be *via* side-chain or *via* backbone. Thus, ligand binding sites are created from primary structure through protein folding. This requirement for *precise* protein folding in the creation of such biologically important local structures is a major reason why they are more effective in assigning function to proteins than primary sequence (no protein folding) and overall global fold (imprecise protein folding and neglect of side chains).

**Figure 2. The Overall Methodology.** The overall methodology may be regarded as being composed of two parts: Part I: the 3D motif consensus and search motif determination (left half), and Part II: the actual application of the screening procedure itself to an application/test set (right half), taking information 'learned' from (Part I). In this diagram, the test/application set in (Part II) is depicted as having been predicted from threading (123D) and modeling (Modeller6v2) as in the present work, but they may also be experimentally solved structures, e.g., from X-ray crystallography and protein NMR.

**Figure 3.** ***All-Atom Representation vs. Double-Centroid Reduced Representation.*** The double-centroid reduced representations (DCRR) is derived from the all-atom representation (AAR). In DCRR, each residue is 'split' into backbone atoms (N, CA, C', O) and side chain atoms (CB, etc.), the centroid of each group computed, and the residue represented by the two centroids. An artificial ligand is shown interacting with residues at its binding site; the interactions in the all-atom representation are H-bonds and VDW attractive forces; in DCRR, they are non-physical interactions, and we call them 'associations.'

**Figure 4.** ***General Structure of the 3D Search Motif.*** The general structure of the 3D search motif is an irregular tetrahedron, with one corner designated as the 'root', R, and three others, the three 'nodes', n1, n2 and n3. As such it is also called a 'tree' data structure. Higher-order irregular polyhedra such as a pentahedron, etc., are possible, but not as computationally efficient for use as a tetrahedron (see text). Edges emanating from the root are called 'branches', while those that join two nodes are designated 'node-edges.' The root and nodes represent amino acid residue centroids (backbone or side chain) in association with ligand atoms.

**Figure 5. Nearest Neighbor Analysis of a Ligand Bound in Its Binding Site in a Protein.** The first step is to find the spheres with centers at each atom in the ligand and with radii typically 4.0 Å; this is done for all structures in the training set. Amino acid atoms in the protein binding site lying within these spheres are taken and further processed. Those which can potentially hydrogen bond or form van der Waals interaction with any ligand atoms (with correct distances and atom identities, see Reyes, V.M. & Sheth, V.N., 2011), are taken and further analyzed to determine which of them are the most common among all the structures in the training set (see next Figure).

**Figure 6. Determination of the 3D Binding Site Motif Consensus from Results of Nearest Neighbor Analysis.** A table is constructed with the training structures as the columns and each ligand atom as the rows; in the figure is a theoretical ligand with 12 atoms, and there are 10 protein structures in the training set. Each cell in the table contains all possible H-bond and VDW pairs between the ligand atoms and amino acid atoms in the binding site of the particular training structure. Those which recur with greatest frequency among the training structures are encircled: red for ligand atom 2, green ligand atom 5, blue for ligand atom 8 and purple for ligand atom 12. These four ligand atoms are then taken to be the 3D binding site motif consensus for this specific ligand; since there are four of them in 3D space, they form a tetrahedron.

**Figure 7 (A, B). The Search Algorithm Illustrated for a Theoretical Ligand and its Protein Receptor.** Panel A shows a theoretical 3D SM with nodes of the tree-structured 3D motif labeled (one root and three branch nodes, branches and node-edges), as well and theoretical distances between each nodes and the fuzzy additive term, $\epsilon$. Panel B shows the 3D SM search algorithm in finer detail. The boxed items in green are Fortran programs (namely, 2, 3a, 3b, 5, 6, 8, 10, 12 and 13) that take input(s) and produce output(s), as indicated by the arrows, and appropriately labeled. Note that the very first input is the protein structure in all-atom representation (AAR), and the very final output is the 3D SM. The eight programs (excluding 13, which is currently done by inspection) have been incorporated together into one script and may be executed in one keystroke.

**Figure 8 (A, B).** ***The 3D Binding Site Motif Consensus for the GTP- and ATP-binding Families.*** The 3D binding site consensus motif for GTP in the small, Ras- type G-protein family (Panel A), and for ATP in the ser/thr protein kinase family (Panel B) are shown. The triphosphate portion of either ligand is not shown in full, as it is disregarded in determining the consensus (as well as in building the 3D search motif). The protein residues engaged in these interactions are drawn in short-hand notation, i.e., two perpendicular arrows intersecting at the midpoint of one (the backbone part) and at the endpoint of the other (the side chain part), which we call the 'backbone arrow' and 'side chain arrow', respectively. The arrowhead of the backbone arrow represents the backbone carbonyl oxygen, its midpoint the C-α atom and its endpoint the backbone amide nitrogen. The arrowhead of the side chain arrow represents either the H-bonding or the terminal atom in the side chain, as the case may be. The one-letter designations of the amino acid residues are indicated; the forward slashes indicate disjunction ("or"). The red dashed lines are H-bonds, and the blue ones VDW interactions, between protein and ligand atoms.

**Figure 9 (A, B).** ***The 3D Search Motifs for the GTP- and ATP-binding Families.*** The 3D search motifs derived from their respective consensus motifs are shown for the GTP-binding families, small Ras-type G-proteins (Panel A) and mRNA capping enzymes (Panel B), and the ATP-binding family ser/thr protein kinases (Panel C). The ligands are shown with their standard atom designations, except for the triphosphate tail, only whose location is shown. The root and three nodes of the search motifs are indicated as R, n1, n2 and n3, respectively. Shown also are the one-letter

designations of the amino acid residues in association with the ligand atoms, the qualitative input parameters of the algorithm. Each is shown in DCRR as an arrowhead connected by a dashed line to a circle; the former represents the residue's side chain centroid, while the latter, the residue's backbone centroid. The lengths of the six sides of the search motif (tetrahedron), which are the quantitative input parameters of the screening algorithm, are shown in the small table at the bottom right of the figure.

**Figure 10.** ***Partial ROC Curves.*** Partial ROC curves deduced from a single operating point on the curve are shown. Such ROC curves are shown for the GTP-binding small, Ras-type G-protein family (Panel A, left), and the ATP-binding ser/thr protein kinase family (Panel B, right). The operating point on each curve is within the high-specificity band; that in panel A (for GTP binding) is also within the high sensitivity band, while that in panel B (for ATP binding) is within the high-middle sensitivity band.

**Figure 11, Validation Screening Results.** Panel A: GTP Binding Site Screening Results from the small Ras-Type G-Protein Family. The Venn Diagram summarizes the results of the screening of the screening of the small ras-type G-protein expanded validation set composed of 45 protein structures. Note that the screening is 100% successful: the algorithm detected GTP- binding site in all the test proteins. Panel B: ATP Binding Site Screening Results from the Protein Kinase Family. The Venn Diagram summarizes the results of the screening of the ser/thr protein kinase expanded validation set composed of 31 protein structures. Note that the algorithm found the ATP binding site in all but two (1CJA and 1NW1) of the 31 structures. Note also that even for the structures that contained no ATP nor ATP analog, the algorithm detected ATP binding sites in all but one (1NW1; right circle).

**TABLE CAPTIONS:**

**Table 1. Panel A.** Training set for GTP-binding proteins.
**Table 1, Panel B.** Training set for ATP-binding proteins.

**Table 2. Panel A.** Positive Controls for GTP-binding proteins. See also page 6 of text for expanded control set.
**Table 2, Panel B.** Positive Controls for ATP-binding proteins. See also page 7 of text for expanded control set.

**Table 3.** Negative Controls for GTP- and ATP-binding proteins.

# FIGURES:

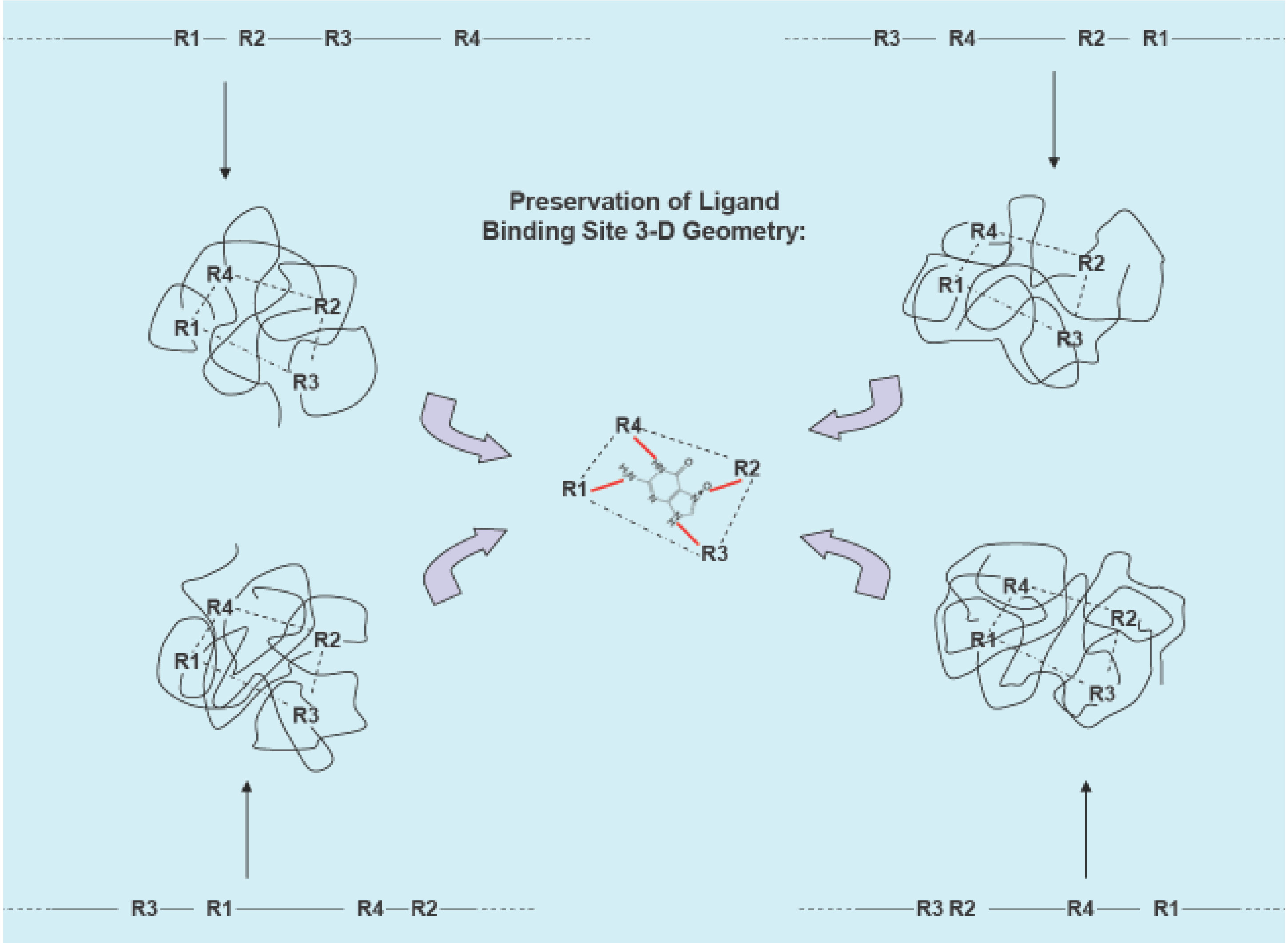

FIGURE 1.

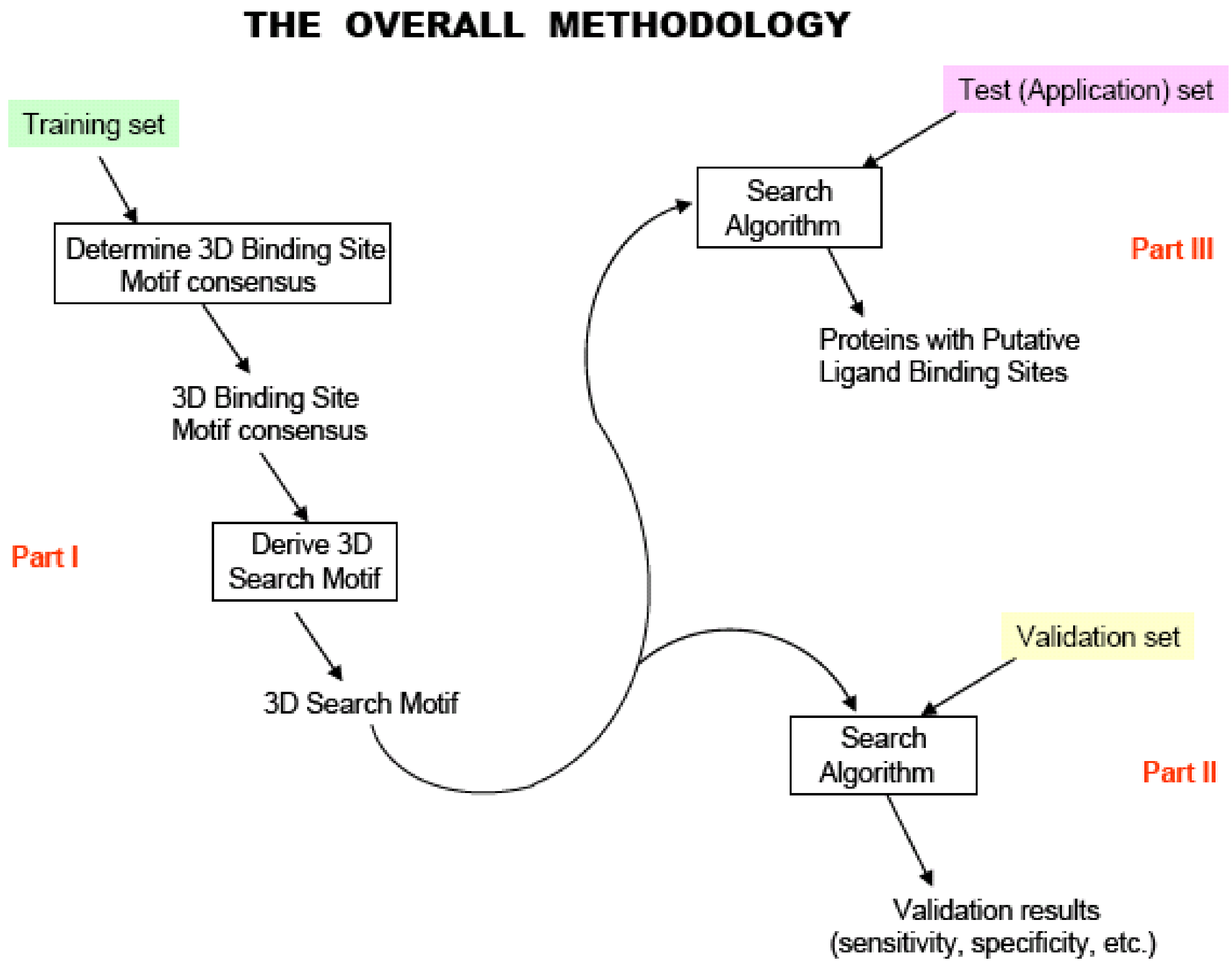
THE OVERALL METHODOLOGY
Training set
Determine 3D Binding Site Motif consensus
3D Binding Site Motif consensus
Derive 3D Search Motif
3D Search Motif
Part I
Test (Application) set
Search Algorithm
Part III
Proteins with Putative Ligand Binding Sites
Validation set
Search Algorithm
Part II
Validation results (sensitivity, specificity, etc.)

**FIGURE 2.**

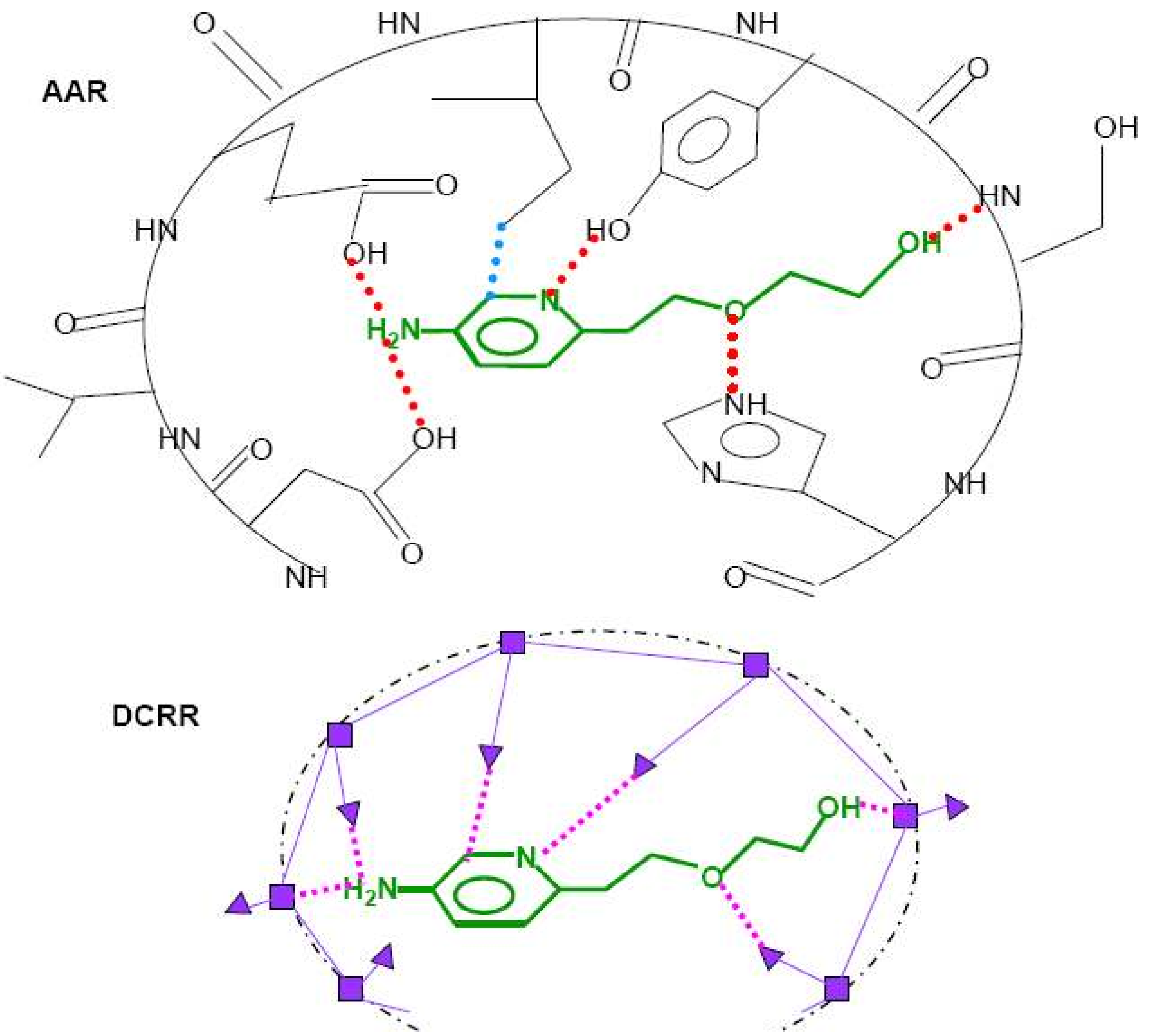

**FIGURE 3.**

**General 3D Search Motif Structure:**

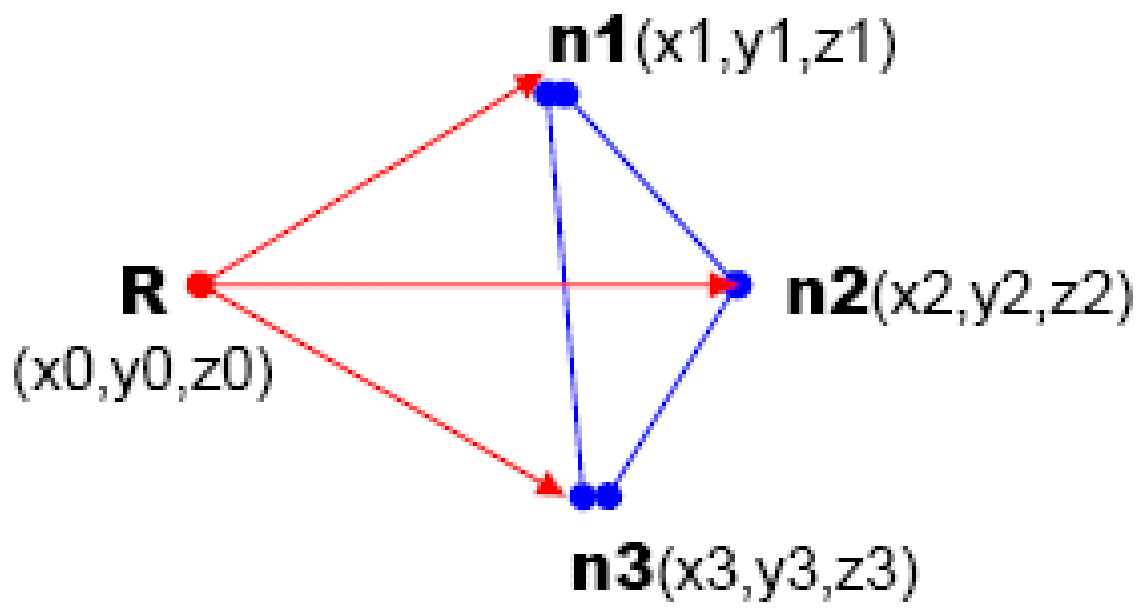

**Screening Algorithm Parameters:**

I. **Qualitative parameters: at least 4; amino acid identities of:**

   **R** == root

   **n1, n2, n3** == nodes

II. **Quantitative parameters: at least 6; lengths of:**

   **Rn1, Rn2, Rn3** == branches

   **n1n2, n2n3, n3n1** == node-edges

III. **Fuzzy factor added to branch and node-edge lengths: ± ε (~1.40 Å)**

**FIGURE 4.**

## The 3D SM Search Procedure

sample
3D SM:

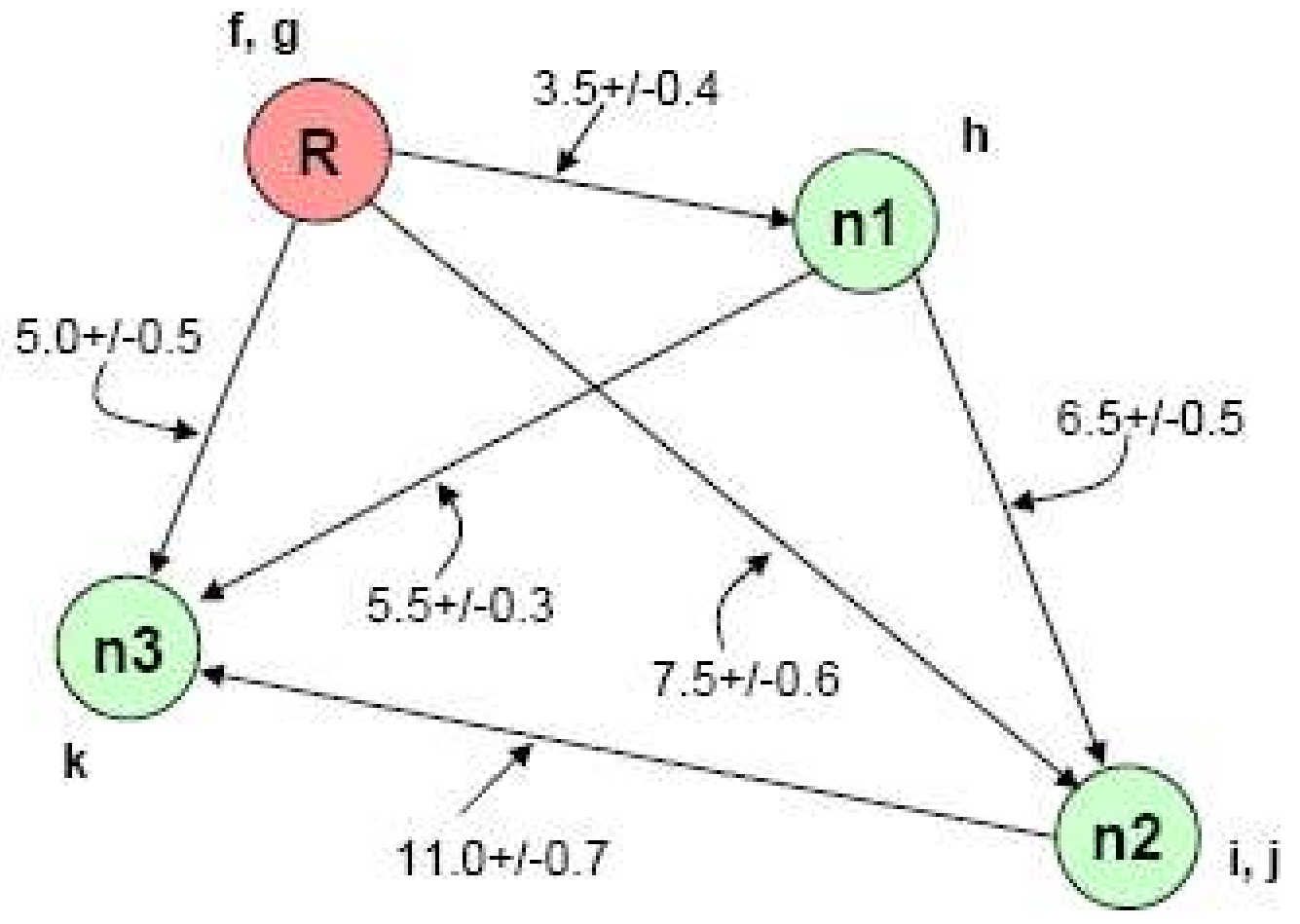

root = **R**

tree structure = **R** → **n1**, **R** → **n2**, **R** → **n3**

nodes = **n1, n2, n3**

branches = **R-n1, R-n2, R-n3**

node-edges = **n1-n2, n2-n3, n1-n3**

**R, n1, n2, n3** = placeholders for consensus residues in the 3D SM

**f, g, h, i, j, k** = identities of amino acid residues in the 3D SM

**x+/-y** == x is average length in Å; y is typically 1.0 standard deviation in Å

**FIGURE 5 A.**

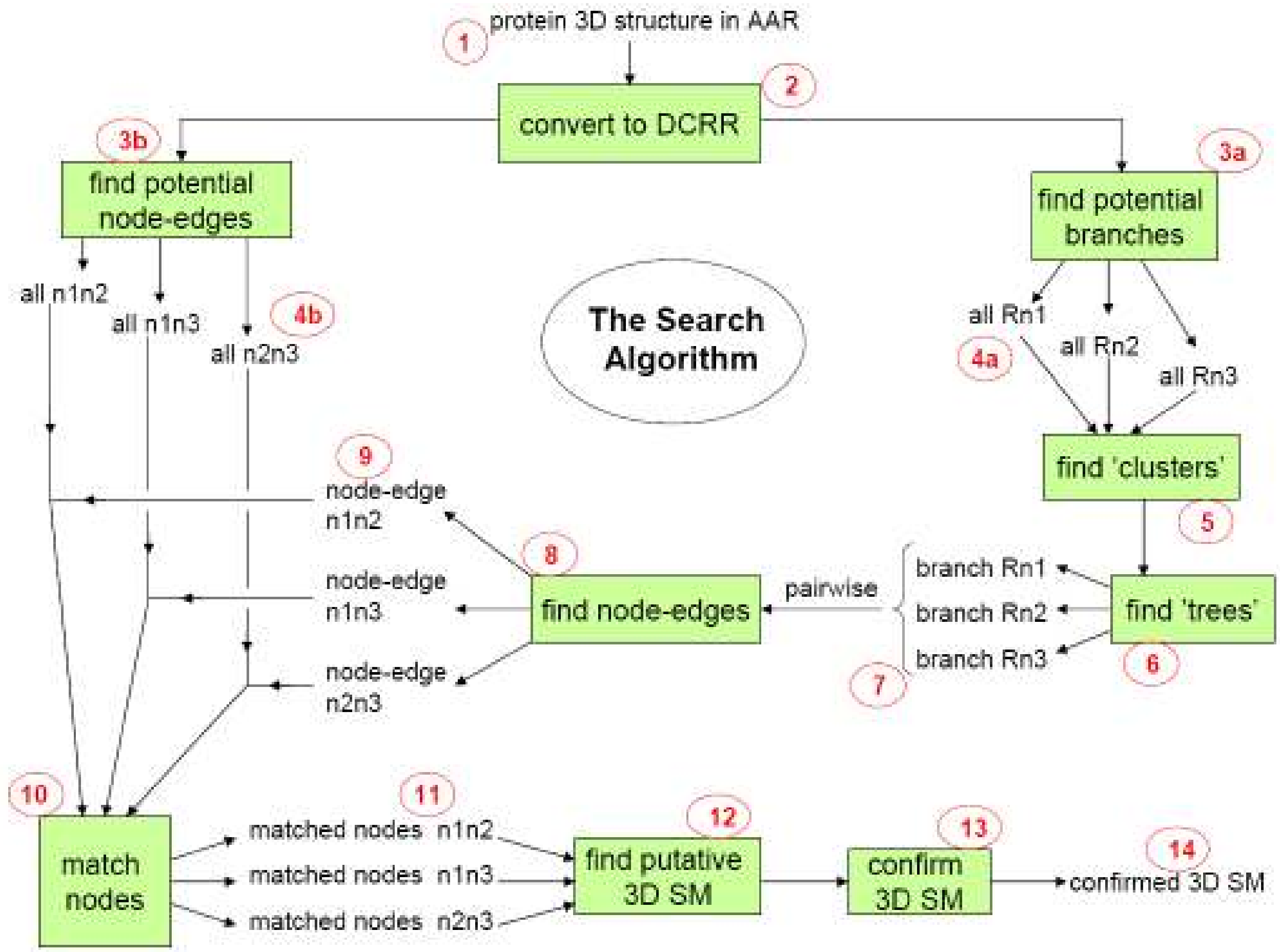

**FIGURE 5 B.**

**H-Bonding and VDW Interactions from Nearest Neighbor Analysis:**

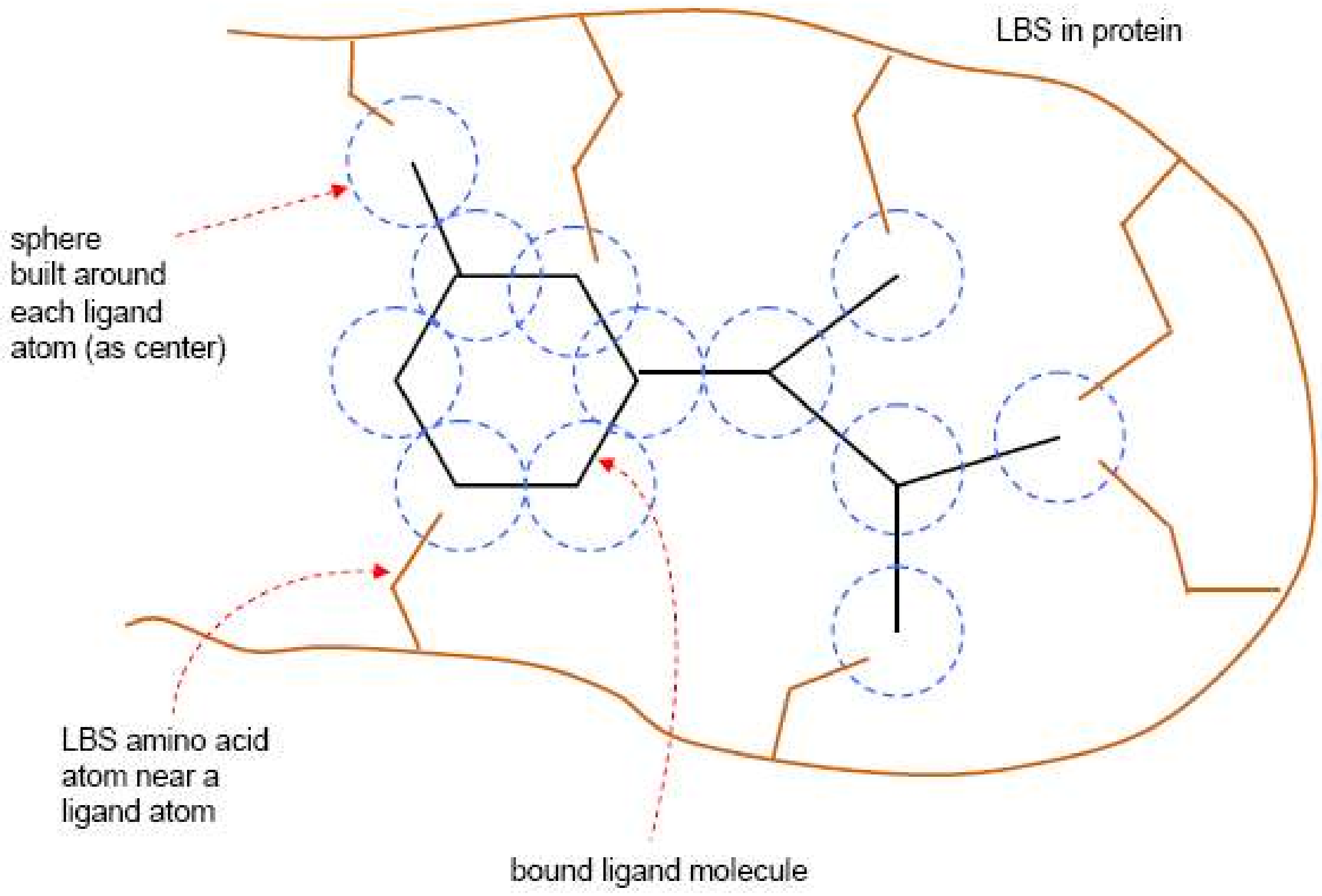

**FIGURE 6.**

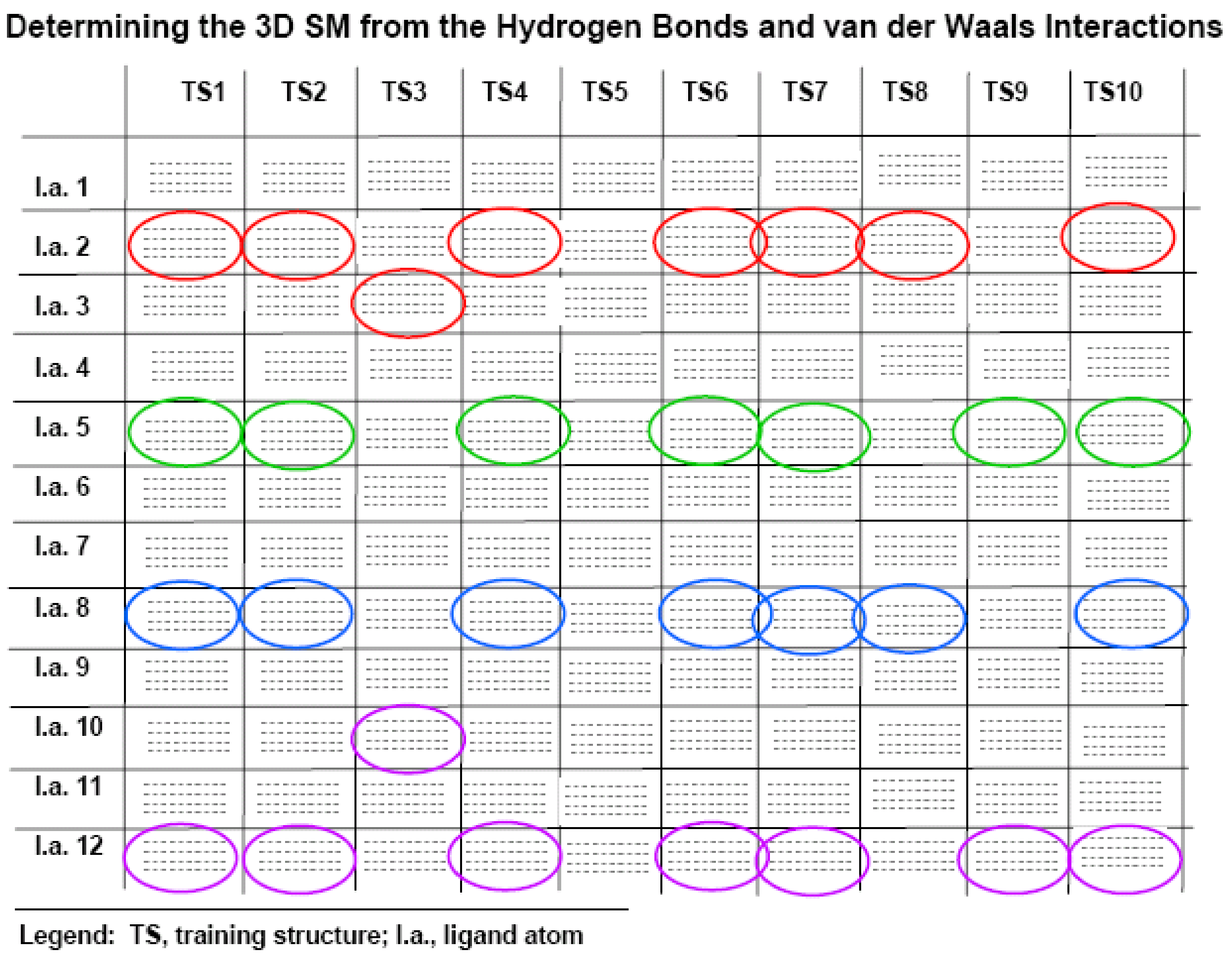
Determining the 3D SM from the Hydrogen Bonds and van der Waals Interactions
TS1
TS2
TS3
TS4
TS5
TS6
TS7
TS8
TS9
TS10
l.a. 1
l.a. 2
l.a. 3
l.a. 4
l.a. 5
l.a. 6
l.a. 7
l.a. 8
l.a. 9
l.a. 10
l.a. 11
l.a. 12
Legend: TS, training structure; l.a., ligand atom

**FIGURE 7.**

**Small, Ras-Type G-Protein GTP Binding Site Consensus Motif**

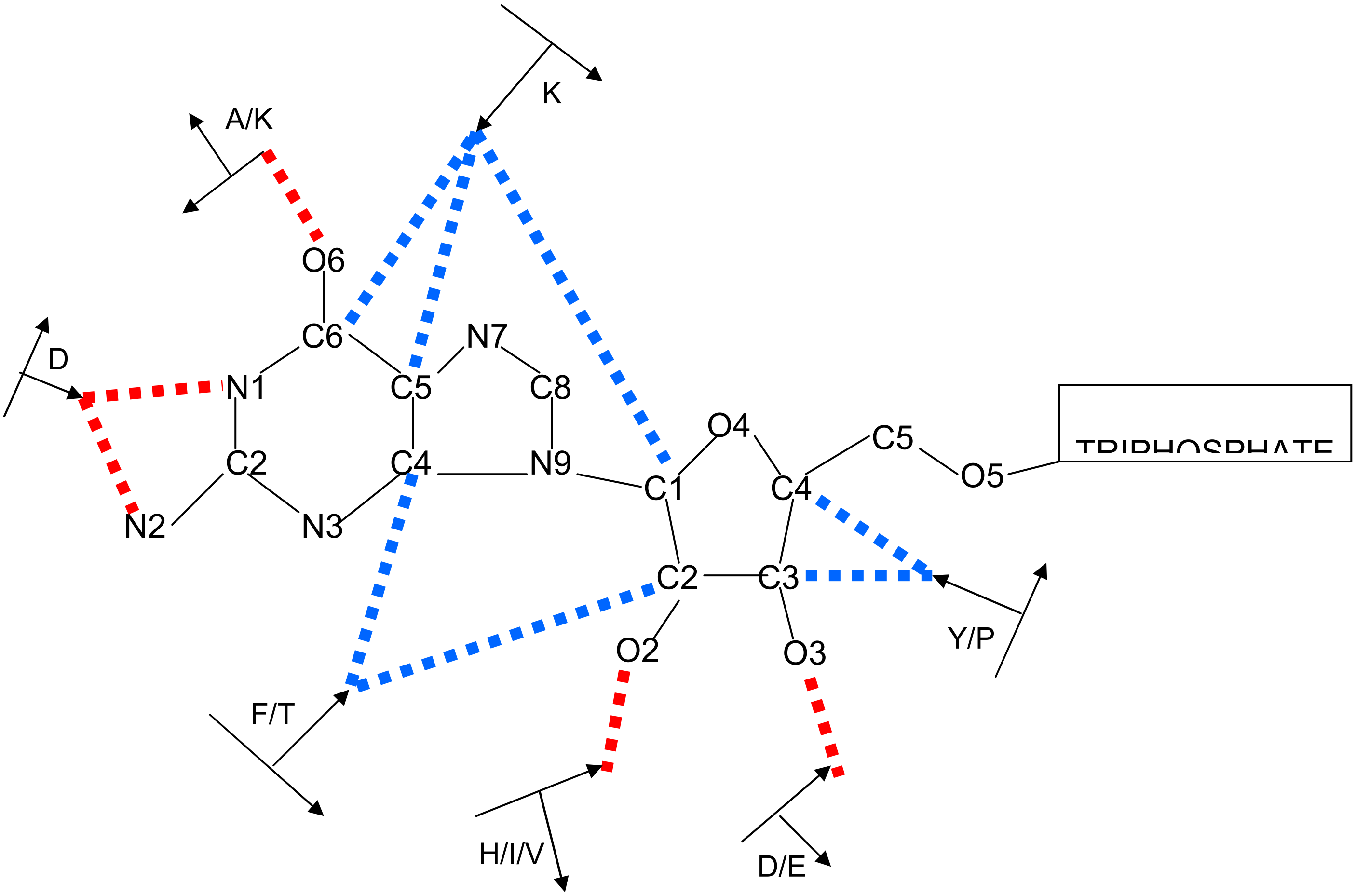

**FIGURE 8 A.**

## Ser/Thr PK ATP Binding Site Consensus Motif

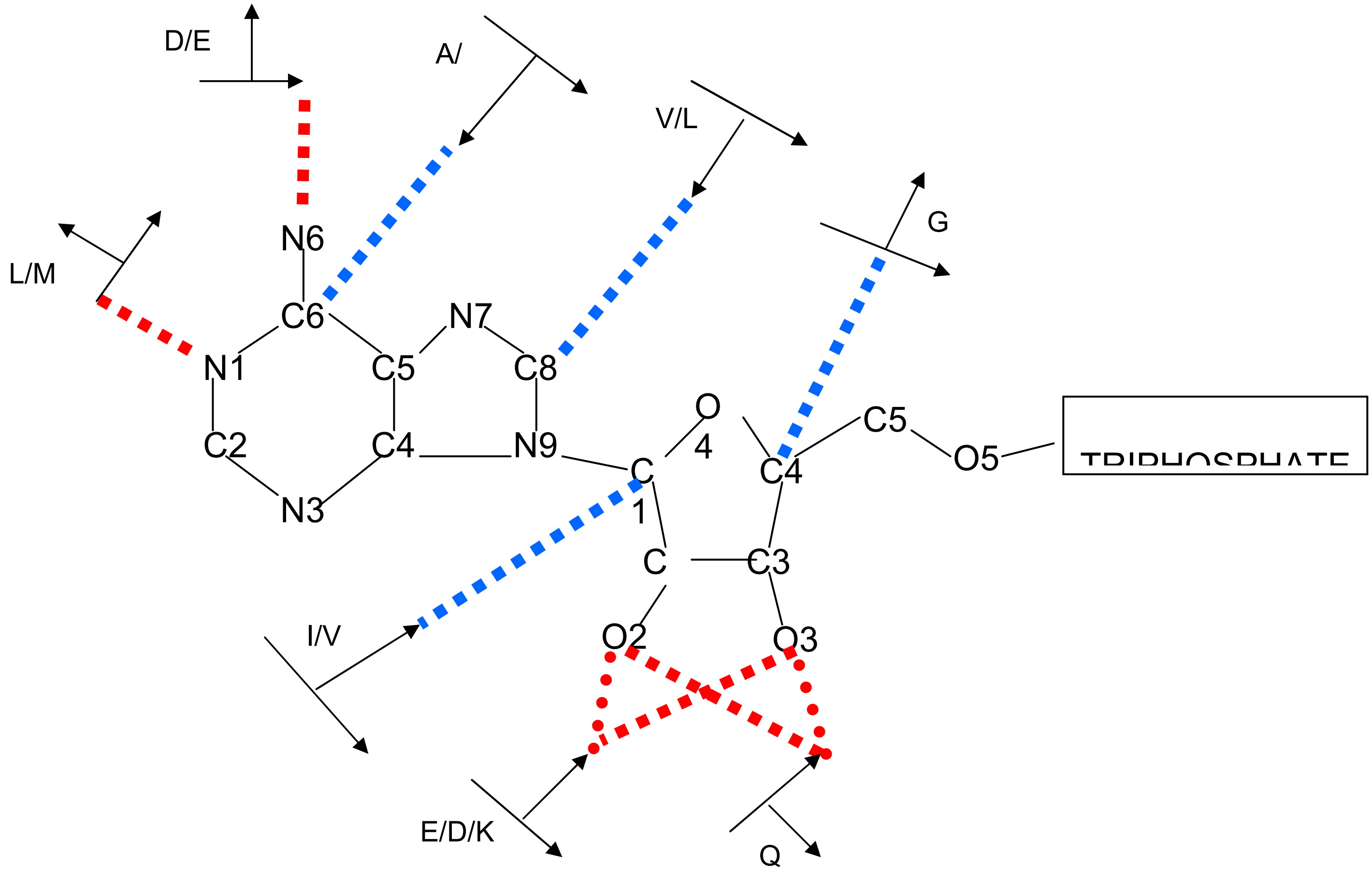

**FIGURE 8 B.**

**GTP molecule (family 02B; small, Ras-type G protein)**

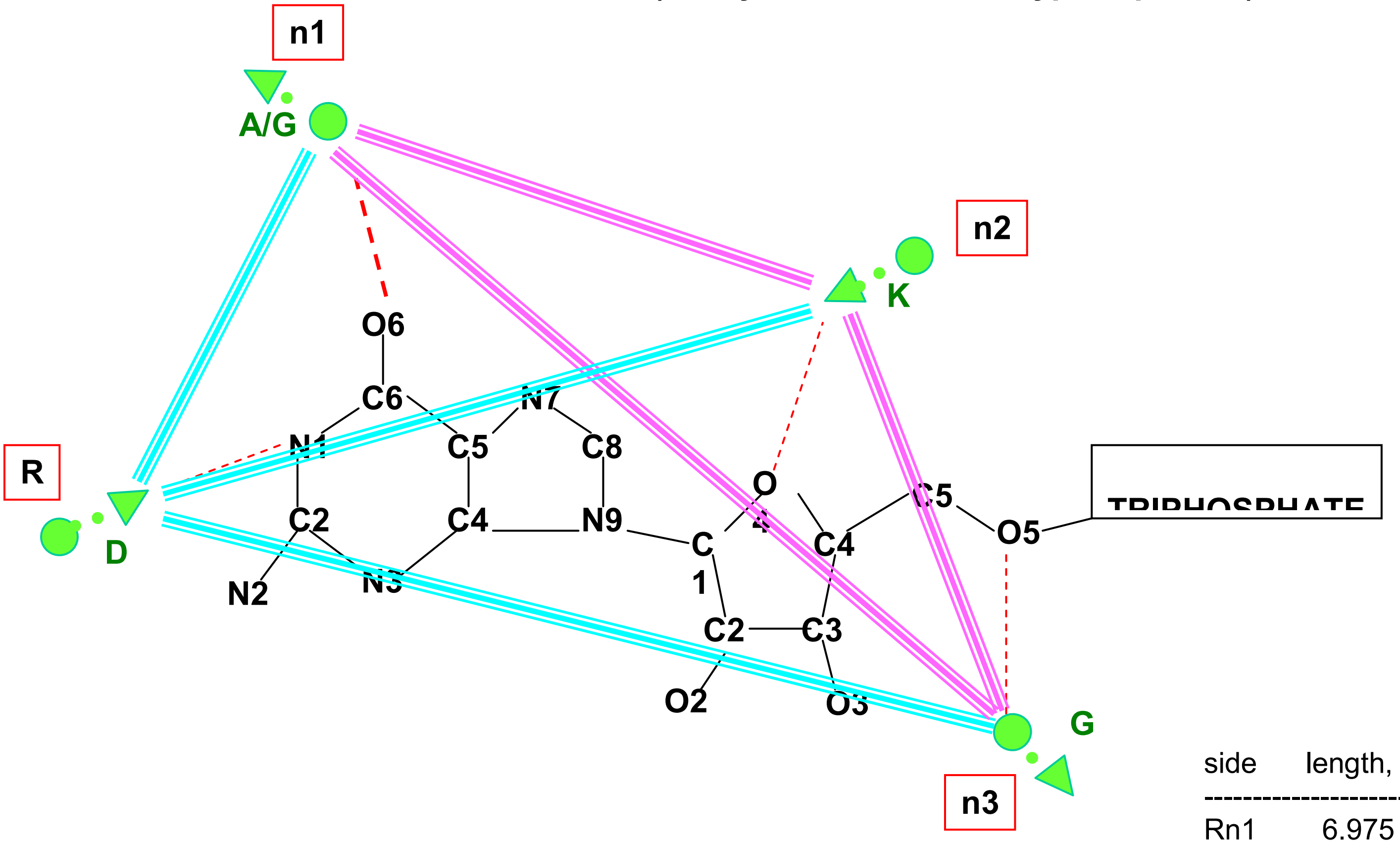

| side | length, Å |
|---|---|
| Rn1 | 6.975 |
| Rn2 | 6.951 |
| Rn3 | 10.732 |
| e1e2 | 8.544 |
| e1e3 | 8.002 |
| e2e3 | 5.650 |

**FIGURE 9 A.**

**ATP molecule (ser/thr PK family)**

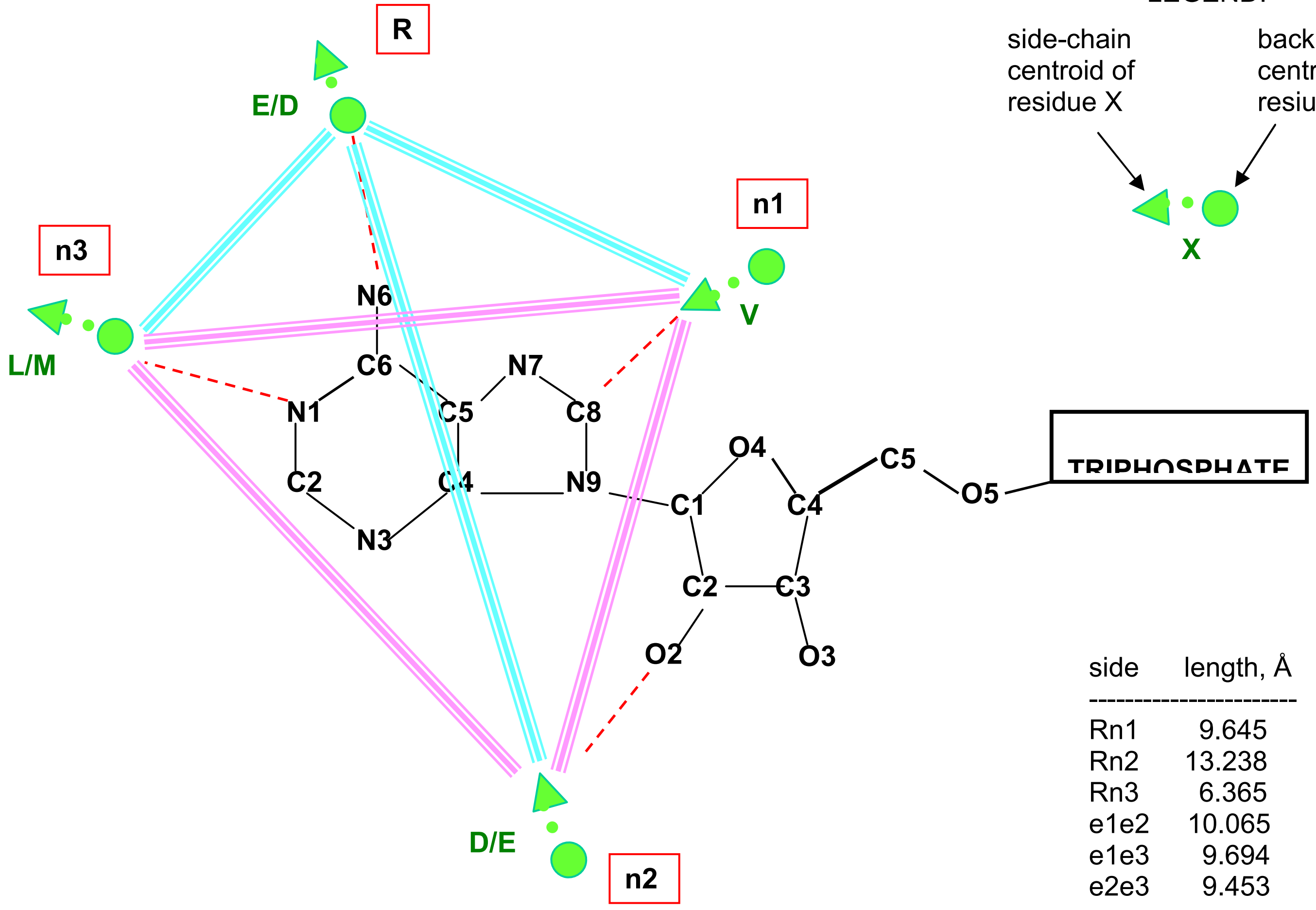

| side | length, Å |
|---|---|
| Rn1 | 9.645 |
| Rn2 | 13.238 |
| Rn3 | 6.365 |
| e1e2 | 10.065 |
| e1e3 | 9.694 |
| e2e3 | 9.453 |

**FIGURE 9 B.**

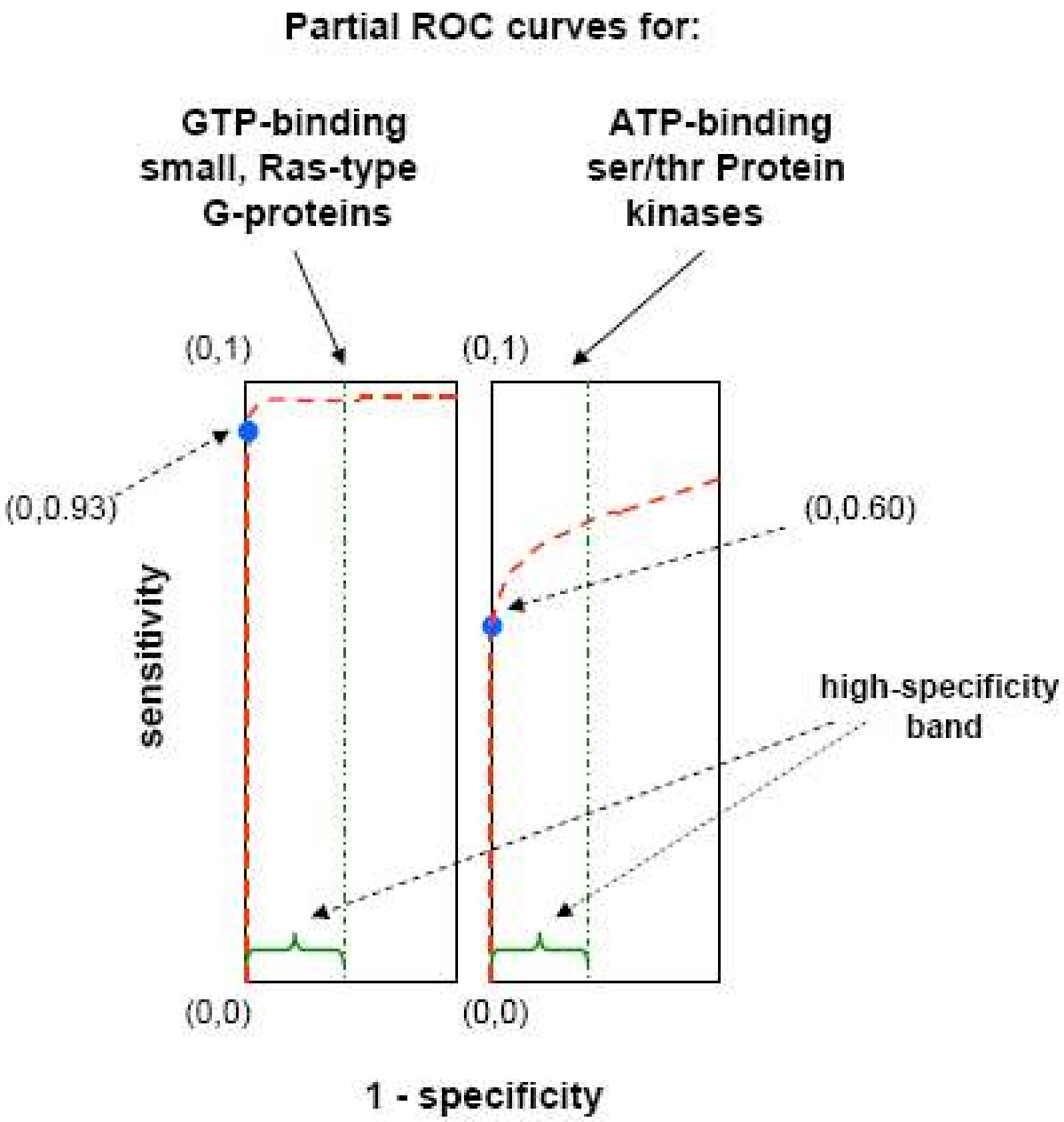

**FIGURE 10, Panels A (right) and B (left).**

Test Set for GTP Binding Site Screening:
45 Ras-type small G-proteins
from the PDB

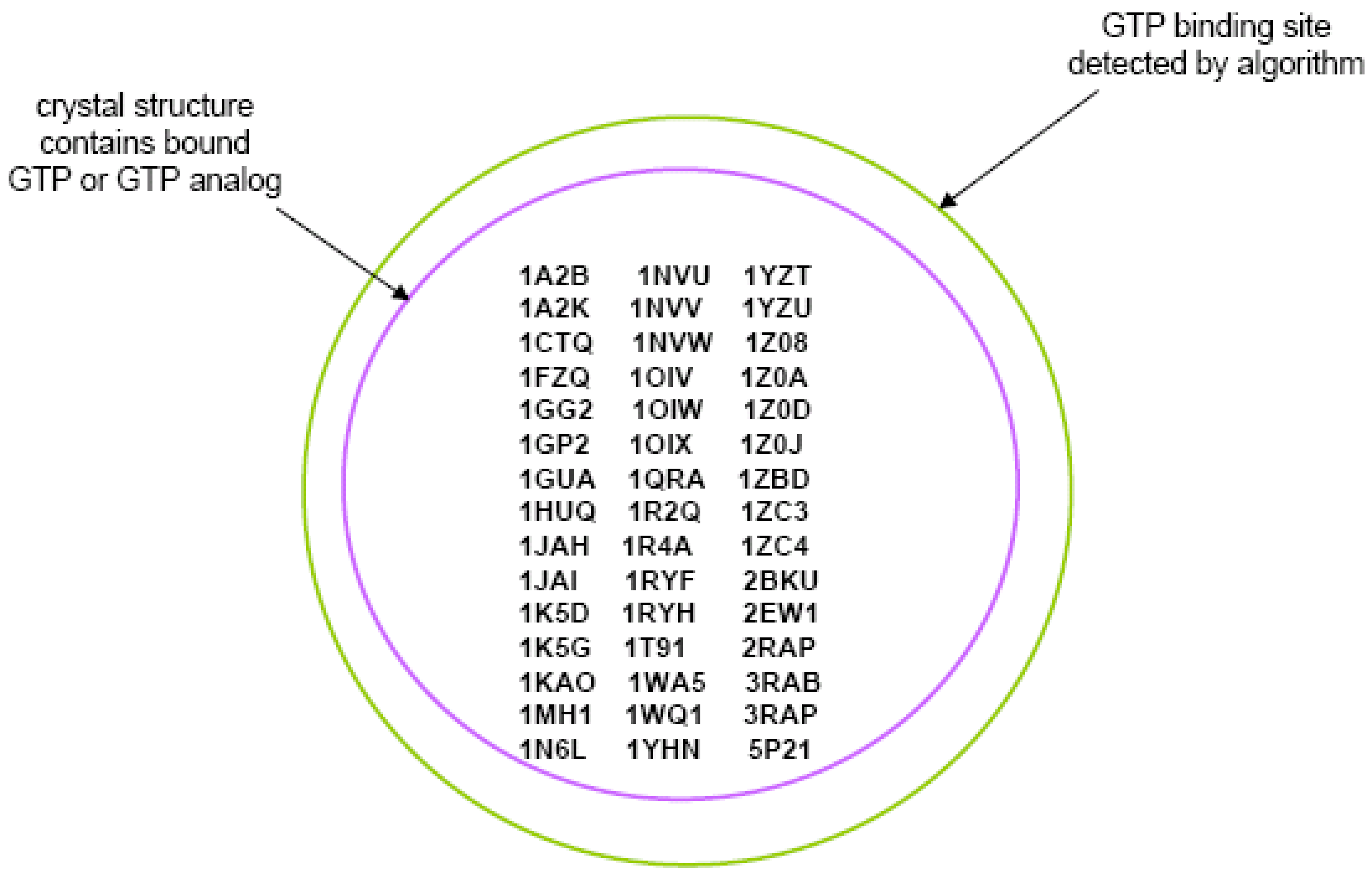

**FIGURE 11 A.**

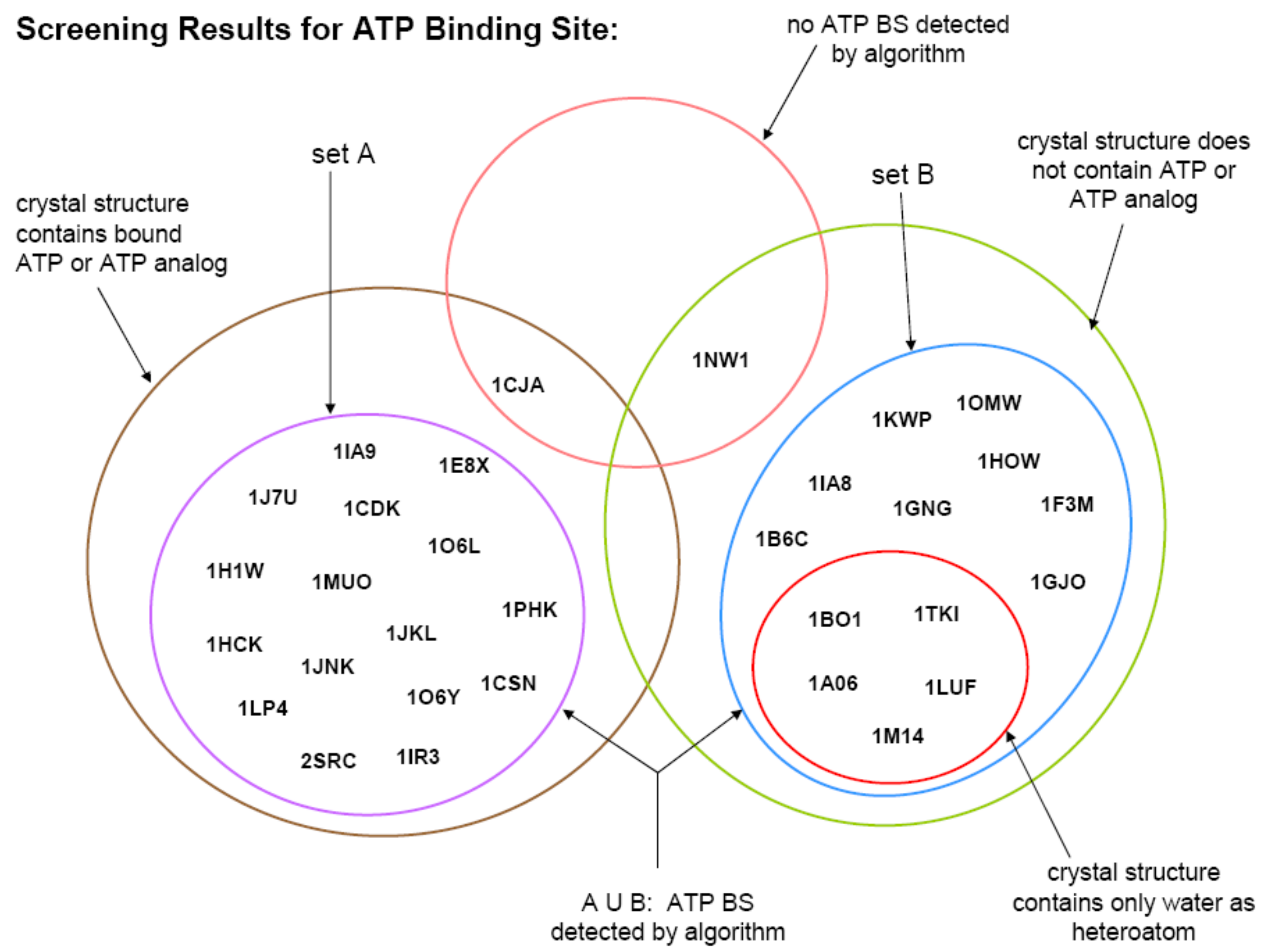
Screening Results for ATP Binding Site:
no ATP BS detected by algorithm
set A
crystal structure contains bound ATP or ATP analog
set B
crystal structure does not contain ATP or ATP analog
1CJA
1NW1
1IA9
1E8X
1J7U
1CDK
1O6L
1H1W
1MUO
1PHK
1HCK
1JKL
1JNK
1O6Y
1CSN
1LP4
2SRC
1IR3
1KWP
1OMW
1HOW
1IA8
1GNG
1F3M
1B6C
1GJO
1BO1
1TKI
1A06
1LUF
1M14
A U B: ATP BS detected by algorithm
crystal structure contains only water as heteroatom

**FIGURE 11 B.**

TABLES:

## Table 1 A: Training Set for GTP-Binding Proteins (families 02B and 02G):

| PDB ID | E.C. No. | Protein Description | Family |
|---|---|---|---|
| 1C4K | 4.1.1.17 | Ornithine Decarboxylase Mutant (Gly121Tyr) | N/A |
| 1E96 | N/A | Structure of RAC/P67Phos Complex | 02B |
| 1FRW | N/A | Structure of E. coli MOBA with Bound GTP and Mn | N/A |
| 1JFF | N/A | Refined Structure of Alpha-Beta Tubulin from Zinc-Induced Sheets Stabilized with Taxol | 02D |
| 1N6L | N/A | Crystal Structure of Human Rab5A A30P Mutant Complexed with GTP | 02B |
| 1NVU | N/A | Structural Evidence for Feedback Activation by Ras-GTP of the Ras-Specific Nucleotide Exchange Factor SOS | 02B |
| 1P16 | 2.7.7.50 | Structure of an mRNA Capping Enzyme Bound to the Phosphorylated Carboxyl Terminal Domain of RNA Polymerase II | 02G |
| 1TUB | N/A | Tubulin Alpha-Beta Dimer, Electron Diffraction | 02D |
| 1A9C | 3.5.4.16 | GTP-Cyclohydrolase I (C110S Mutant) in Complex w/ GTP | N/A |
| 1CKM | 2.7.7.50 | Structure of 2 Different Conformations of mRNA Capping Enzyme in Complex with GTP | 02G |
| 1HWX | 1.4.1.3 | Crystal Structure of Bovine Liver Glutamate Dehydrogenase Complexed w/ GTP, NADH & L-Glu | 02K |
| 1HWZ | 1.4.1.3 | Bovine Glutamate Dehydrogenase Complexed with NADPH, Glutamate and GTP | 02K |
| 1LOO | 6.3.4.4 | Crystal Structure of the Mouse Muscle Adenylo-succinate Synthetase Ligated with GTP | 02B |
| 1M7B | N/A | Crystal Structure of RND3/RHOE: Functional Implications | 02B |
| 1O3Y | N/A | Crystal Structure of Mouse ARF1(Delta17-Q71L), GTP Form | 02B |
| 2RAP | N/A | The Small G-Protein RAP2A in Complex with GTP | 02B |

## Table 1 B: Training Set for ATP-Binding Proteins (family 01a):

| PDB ID | E.C. No. | Protein Description | Kinase Type |
|---|---|---|---|
| 1B38 | 2.7.1.37 | Human Cyclin-Dependent Kinase 2 | CDK2 |
| 1B39 | 2.7.1.37 | Human Cyclin-Dependent Kinase 2 Phosphorylated on Thr 160 | CDK2 |
| 1FIN | 2.7.1.- | Cyclin A-Cyclin-Dependent Kinase 2 Complex | CDK2 |
| 1GOL | 2.7.1.- | Rat MAP Kinase ERK2 with an Arg Mutation at Position 52 | MAPK |
| 1HCK | 2.7.1.37 | Human Cyclin-Dependent Kinase 2 | CDK2 |
| 1JST | 2.7.1.- | Phosphorylated Cyclin-Dependent Kinase 2 Bound to Cyclin A | CDK2 |
| 1PHK | 2.7.1.38 | Two Structures of the Catalytic Domain of Phoshorylase Kinase: An Active Protein Kinase Complexed with Nucleotide, Substrate-Analogue and Product | Phos. Kin. |
| 1QL6 | 2.7.1.38 | The Catalytic Mechanism of Phosphorylase Kinase Probed by Mutational Studies | Phos. Kin. |
| 1QMZ | 2.7.1.- | Phosphorylated CDK2-Cyclin A-Substrate Peptide Complex | CDK2 |
| 2PHK | 2.7.1.38 | The Crystal Structure of a Phosphorylase Kinase Peptide-Substrate Complex: Kinase Substrate Recognition | Phos. Kin. |

**Table 1 A,B. The Training Structures for the GTP- and ATP-binding Families**

| PDB ID | DESCRIPTION |
|---|---|
| | **Positive Controls for the GTP-Binding small Ras-type G-protein Family** |
| **1C1Y** | Crystal Structure of RAP.Gmppnp in Complex with the Ras-Binding-Domain of C-RAF1 Kinase (RAFRBD) |
| **1GWN** | The Crystal Structure of the Core Domain of Rhoe/Rnd3 - A Constitutively Activated Small G Protein |
| **1NVX** | Structural Evidence For Feedback Activation by Rasgtp of the Ras-Specific Nucleotide Exchange Factor SOS |
| **1PLK** | C-H-Ras p21 Protein Mutant with Gly 12 Replaced by Pro (G12P) Complexed with Guanosine-Triphosphate; Protein from Human (Homo Sapiens) Cellular Harvey-Ras Gene Truncated and Expressed in Escherichia Coli |
| **1QRA** | Structure of p21Ras in Complex with GTP at 100 K |
| **1ZBD** | Structural Basis of RAB Effector Specificity: Crystal Structure of the Small G Protein RAB3a Complexed with the Effector Domain of Rabphilin-3a |
| **3RAP** | The Small G Protein RAP2 in a Non-Catalytic Complex with GTP |
| **521P** | H-Ras P21 Protein Mutant with Gly 12 Replaced by Val (G12V) Complexed with Guanosine Triphosphate; Protein from Human (Homo Sapiens) Cellular Harvey-Ras Gene Truncated and Expressed in Escherichia Coli |
| **1AS3** | G42V Mutant Form of Rat GIA1 Complexed with GDP |
| **1BOF** | Rat GIA1 Bound to GDP and Magnesium |
| **1GIT** | G203A Mutant Form of Rat GIA1 Bound with Phosphate and Mg++ |
| **1KAO** | Human RAP2a Complexed with GTP, GDP and GTP-gamma-S |
| **1PLL** | G12P Mutant Form of Human H-Ras p21 Protein Complexed with GDP |
| **1TAG** | Bovine Transducin Alpha Complexed with GDP and Mg++ |
| **2RAP** | Human RAP2a Protein in Complex with GTP |

**Table 2 A.**

| Positive Controls for the ATP-Binding ser/thr Protein Kinase Family | |
|---|---|
| **PDB ID** | **DESCRIPTION** |
| **1CDK** | cAMP-Dependent Protein Kinase Catalytic Subunit (E.C.2.7.1.37) (Protein Kinase A) Complexed with Protein Kinase Inhibitor Peptide Fragment 5-24 (Pki(5-24) Isoelectric Variant Ca) and Mn2+ Adenylyl Imidodiphosphate (MnAMP-PNP) at pH 5.6 and 7c and 4c |
| **1FMO** | Crystal Structure of a Polyhistidine-Tagged Recombinant Catalytic Subunit Of cAMP-Dependent Protein Kinase Complexed with the Peptide Inhibitor PKI(5-24) and Adenosine |
| **1GY3** | PCDK2/Cyclin A in Complex With MgADP, Nitrate and Peptide Substrate |
| **1JBP** | Crystal Structure of the Catalytic Subunit of cAMP-Dependent Protein Kinase Complexed with a Substrate Peptide, ADP and Detergent |
| **1Q24** | PKA Double Mutant Model of PKB in Complex with MgATP |
| **1S9I** | X-Ray Structure of the Human Mitogen-Activated Protein Kinase Kinase 2 (MEK2) in a Complex with Ligand and MgATP |
| **1S9J** | X-Ray Structure of the Human Mitogen-Activated Protein Kinase Kinase 1 (ME1) in a Complex with Ligand and MgATP |
| **1UA2** | Crystal Structure of Human CDK7 |
| **2CPK** | cAMP-Dependent Protein Kinase (E.C.2.7.1.37) (cAPK) (Catalytic Subunit) from Recombinant Mouse (Mus Musculus) "Alpha" Isoenzyme Expressed in Escherichia Coli |
| **1CSN** | Binary Complex of Casein Kinase-1 from S. Pombe with Mg-ATP |
| **1H1W** | Crystal Structure of Human PDK1 (Phosphoinositide-Dependent Kinase-1) Catalytic Domain |
| **1L3R** | Crystal Structure of A Transition State Mimic of the Catalytic Subunit of cAMP-Dependent Protein Kinase |
| **1OGU** | Structure Of Human Thr160-Phospho CDK2/Cyclin A Complexed with A 2-arylamino-4-cyclohexylmethyl-5-nitroso-6-aminopyrimidine Inhibitor |
| **1RDQ** | Hydrolysis of ATP in the Crystal of Y204A Mutant of cAMP-Dependent Protein Kinase |
| **1CM8** | Phosphorylated Human MAP Kinase P38-Gamma |

**Table 2 B.**

| PDB ID | DESCRIPTION |
|---|---|
| 104M | Sperm whale (*Physeter catodon*) skeletal muscle myoglobin (heme-iron[II]-bound) with bound N-butyl isocyanide and sulfate ion at pH 7.0 |
| 1ASH | Iron(II)-protoporphyrin IX-bound hemoglobin domain I from *Ascaris suum* with bound dioxygen at 2.2 Å resolution |
| 1B3B | Structure of glutamate dehydrogenase from *Thermotoga maritima* with mutations N97D and G376K |
| 1BRF | Structure of Rubredoxin with bound Fe(III) from Pyrococcus furiosus at 0.95 Å resolution |
| 1CBN | Structure of the hydrophobic protein crambin from the seed of *Crambe abyssinica* (Abyssinian cabbage) at 130°K and at 0.83 Å resolution |
| 1CKO | Structure of mRNA capping enzyme from *Chlorella* virus PBCV-1 in complex with the CAP analog GpppG |
| 1CRP(1) | NMR structure (n=20) of human C-H-Ras p21 protein (catalytic domain, res. 1-166) complexed with GDP and MG |
| 1EWK(1) | Structure of the metabotropic glutamate receptor subtype 1 from *Rattus norvegicus* complexed with glutamate |
| 1F3O | Structure of MJ0796 ATP-binding cassette with bound Mg-ADP from *Methanococcus jannaschii* |
| 1FW5 | Solution structure of membrane binding peptide of Semliki forest virus mRNA capping enzyme NSP1 |
| 1HWY | Glutamate dehydrogenase from *Bos taurus* complexed with NAD and 2-oxoglutarate |
| 1JBP(2) | Structure of the catalytic subunit of mouse cAMP-dependent protein kinase complexed with a substrate peptide, ADP and detergent |
| 1JFF | Refined structure of bovine (Bos taurus) α-β tubulin from zinc-induced sheets stabilized with taxol |
| 1MJJ | Structure of the complex of the Fab fragment of esterolytic antibody MS6-12 and the transition-state analog, N-{[2-({[1-(4-carboxybutanoyl)amino]-2-phenylethyl}-hydroxyphosphinyl)oxy]acetyl}-2-phenylethylamine |
| 1MV5 | Structure of the ATP-binding domain of the multidrug resistance ABC transporter and permease protein from *Lactococcus lactis* with bound ADP, ATP and Mg ion |
| 1NQT | Structure of glutamate dehydrogenase from *Bos taurus* with bound ADP |
| 1OGU(2) | Structure of human Thr160-phospho CDK2/cyclin A complexed with a 2-arylamino-4-cyclohexylmethyl-5-nitroso-6-aminopyrimidine inhibitor |
| 1PE6 | Structure of papain (E.C.4.3.22.2) from the papaya fruit (*Carica papaya*) latex complexed with E-64-C ((2S,3S)-3-(1-(N-(3-methylbutyl)amino)-leucylcarboxyl)oxirane-2-carboxylate) at 2.1 Å resolution |
| 1RDQ(2) | Hydrolysis of ATP in the crystal of Y204A mutant of cAMP-dependent protein kinase from *Mus musculus* |
| 1RQ7 | *Mycobacterium tuberculosis* FTSZ (filamenting temperature-sensitive mutant Z) in complex with GDP |
| 1SVS(1) | Structure of the K180P mutant of GI α subunit bound to GPPNHP (phosphoaminophosphonic acid-guanylate ester) |
| 1TUB | Electron diffraction structrue of *Sus scrofa* (pig) tubulin α-β dimer with bound GTP, GDP and taxotere |
| 1TWY | Structure of a hypothetical ABC-type phosphate transporter from *Vibrio cholerae* O1 Biovar eltor |
| 1Z3C | mRNA cap (guanine-N7) methyltransferasein from the encephalitozooan *Cuniculi* complexed with AzoAdoMet |

(1) used for for ATP (ser/thr protein kinases), but not for GTP (small, Ras-type G-proteins)
(2) used for for GTP (small, Ras-type G-proteins), but not for ATP (ser/thr protein kinases)

**Table 3. The Negative Control Structures for the GTP- and ATP-binding Families**